\documentclass[preprint2]{aastex}
\usepackage{amsfonts}
\usepackage{amssymb}
\usepackage{subfigure}
\usepackage{graphicx,color}
\usepackage{lineno}

\usepackage{ulem}
\usepackage{amsmath}

\newcommand{\Mpc}{\ensuremath{\,{\rm Mpc}}}
\newcommand{\K}{\ensuremath{\, {\rm K}}}
\newcommand{\mK}{\ensuremath{\, {\rm mK}}}

\newcommand{\MHz}{\ensuremath{\, {\rm MHz}}}

\newcommand{\Msun}{\ensuremath{{M_\sun}}}

\begin{document}

\title{{\LARGE Maximum Absorption of the Global 21 cm Spectrum in the Standard Cosmological Model}}
\renewcommand{\thefootnote}{\fnsymbol{footnote}}
\author{\large Yidong Xu\altaffilmark{1}, Bin Yue\altaffilmark{1}, Xuelei Chen\altaffilmark{1,2,3,}\footnote{Corresponding author: Xuelei Chen (xuelei@cosmology.bao.ac.cn)}\\ }
\altaffiltext{1}{National Astronomical Observatories, Chinese Academy of Sciences, Beijing 100101, China}
\altaffiltext{2}{School of Astronomy and Space Science, University of Chinese Academy of Sciences, Beijing 100049, China}
\altaffiltext{3}{Center for High Energy Physics, Peking University, Beijing 100871, China}

\begin{abstract}
The absorption feature in the global spectrum is likely the first 21 cm observable from the cosmic dawn, which
provides valuable insights into the earliest history of structure formation. We run a set of high-resolution hydrodynamic
simulations of early structure formation to assess the effect of non-linear structure formation on the 
maximum absorption level (i.e. assuming the spin temperature coupling is saturated) of the global 21 cm spectrum in the standard cosmological framework. We ignore the star formation and feedbacks, which also tends to reduce the absorption signal,  
but take into account the inevitable non-linear density fluctuations in the intergalactic medium (IGM), 
shock heating, and Compton heating, which can reduce the absorption level. We found that the combination of these 
reduced the maximum absorption signal by $\sim 15\%$ at redshift 17, as compared with the homogeneous or 
linearly fluctuating IGM. These effects have to be carefully accounted for when interpreting the observational results, especially when 
considering the necessity of introducing new physics. 
\end{abstract}

\keywords{Cosmology: theory --- dark ages, reionization, first stars --- 
intergalactic medium}

\maketitle

\section{Introduction}\label{Intro}

The spectrum of the sky-averaged 21 cm brightness temperature, or the so-called global 21 cm signal, 
provides valuable information on the early history of structure formation, from the dark ages to the cosmic reionization
(e.g. \citealt{2012RPPh...75h6901P}). 
The EDGES experiment has reported the detection of an absorption trough with a depth of 
$\delta T_{\rm 21} = -500^{+200}_{-500}\, {\rm mK}$ (99\% confidence level) corresponding to the cosmic dawn
\citep{2018Natur.555...67B}, which is unexpectedly large as compared with theoretical predictions from the standard model. 
Although the claimed signal may be affected by instrumental effects (e.g. \citealt{2019ApJ...874..153B}), 
mis-modeled foregrounds (e.g. \citealt{2018Natur.564E..32H}), or data analysis systematics (e.g. \citealt{2019ApJ...880...26S}), 
various theoretical works have been trying to explain the signal level by introducing new physics.
These models invoke a variety of mechanisms to explain the large absorption, e.g. extra-cooling of the cosmic gas (e.g. 
\citealt{2018Natur.555...71B}; \citealt{2018arXiv180210094M}; \citealt{2018arXiv180210577F};
\citealt{2018arXiv180303091B}; \citealt{2018arXiv180309734S}; \citealt{2018arXiv180310671H};
\citealt{2018arXiv180401092M}; \citealt{houston2018natural}; \citealt{li2018detailed}), extra-source of early radio 
background in addition to the cosmic microwave background (CMB) (e.g. \citealt{2018arXiv180207432F,
2018ApJ...868...63E,2018arXiv180303245F}), or modified Hubble expansion rate  \citep{costa2018interacting,wang2018constraining}.

Many other experiments with a variety of designs are also trying to measure the global 21 cm signal,
such as the SARAS \citep{2018ExA....45..269S}, PRIZM \citep{2019JAI.....850004P}, SCI-HI \citep{2014ApJ...782L...9V}, 
BIGHORNS \citep{2015PASA...32....4S}, LEDA \citep{2018MNRAS.478.4193P}, 
and ASSASSIN \citep{2020MNRAS.499...52M} from ground, and the planed 
DAPPER \citep{2019AAS...23421202B} and DSL \citep{2020arXiv200715794C} from space. 
In particular, the SARAS-2 has already put some constraints on the
21 cm spectrum, and disfavors models that feature weak X-ray heating along with
rapid reionization \citep{2018ApJ...858...54S}.

To correctly interpret the observations,  it is important to  calculate the 21 cm absorption level precisely, taken into account of various effects. 
Although the absorption feature  in the global 21 cm spectrum is produced by the gas which is still quite homogeneous during the cosmic dawn, where the volume fraction of collapsed halos are still small, the budding inhomogeneity can still affect the result. \citet{2018ApJ...869...42X} investigated the effect of gas inhomogeneity on the maximum absorption level of the global 21 cm spectrum. It was found that the non-linearity of the gas 
density fluctuations induced by the structure formation, and the associated adiabatic heating, suppress the signal level  significantly.

Note that the 21 cm absorption feature is produced 
by the neutral hydrogen with spin temperature lower than the CMB temperature at that epoch \citep{chen2004spin,chen200821}.
The neutral hydrogen spin temperature  would generally fall somewhere between the gas kinetic temperature and CMB temperature, depending on the intensity of the Lyman alpha (Ly-$\alpha$) background, which couples the spin and kinetic temperature of the gas.
By maximum, we are referring to the case that the spin-kinetic coupling of the gas
is saturated, such that the spin temperature is equal to the kinetic temperature,  and the largest amount of absorption is produced. 
In realistic models, by the time a strong Ly-$\alpha$ background is set up by star formation and black hole accretion, some 
amount of ionization and heating would have already taken place. The ionized gas would not contribute to the 21 cm signal, while the neutral gas heated above the CMB temperature would appear in 21 cm emission, reducing the total amount of 21 cm absorption. The 
ionization and radiation induced heating would however  depend on many modeling details, which results in a variety of spectrums  \citep{2017MNRAS.472.1915C}. However, to assess the necessity of new physics, we can focus on the maximum absorption case.
As all of these effects reduce the absorption, one can obtain a very conservative limit if we ignore them. 

However, the analytic estimates in \citet{2018ApJ...869...42X} should only be taken qualitatively, as the large-scale
clustering of halos and the structure formation shocks that are inevitable during the cosmic dawn are not easy to model analytically.  
Also, the adopted density profile for the intergalactic medium (IGM) around collapsed halos from the ``infall model'' only applies to density 
peaks \citep{2004MNRAS.347...59B}. Applying it to the whole IGM of any environments requires an artificial normalization, 
which may result in an inaccurate level of gas density fluctuations and the resultant 21 cm signal.
Recently, \citet{2020PhRvD.101h3502V} has also investigated analytically
the maximum amplitude of the high-redshift 21 cm absorption feature,
accounting for 21 cm heating, Lyman-$\alpha$ heating, and the density fluctuations.
Adopting the non-linear density distribution of the MHR00 model \citep{2000ApJ...530....1M}, they found that
the density fluctuations result in a decrement in the absolute value of $\sim 10\%$ in the maximum 21 cm absorption.

In this work, we focus on the maximum signal level of 21 cm brightness from cosmic dawn within the 
standard framework, i.e. assuming no extra cooling or extra radio background from new physics, a fully neutral
IGM before reionization, and saturated coupling between the spin temperature of neutral hydrogen and the kinetic 
temperature of gas. However, we do consider the inevitable standard model evolutions, such as non-linear structure formation and 
Compton heating. To incorporate more reliable density profiles of gas in the IGM, for both over-dense and under-dense 
regions, and to include the clustering effect of non-linear structures, we use a set of high-resolution hydrodynamic 
simulations to compute the expected global 21 cm signal at high redshifts, and discuss various effects that impact the signal
level. 

This paper is organized as follows. We describe our simulations set and present some basic results in section~\ref{global_signal}, and then we discuss the effects of density profiles, the shock heating and Compton heating, and the large-scale clustering, in section~\ref{effects}. 
We conclude in section~\ref{conclusions}. Throughout this paper, we adopt the $\Lambda$CDM model with the 
Planck 2018 cosmological parameters \citep{Planck2018}.

\section{Simulation and the maximum signal}
\label{global_signal}
We use hydrodynamical simulations to investigate the the effects of non-linear structure formation on the global 21 cm signal from 
cosmic dawn. In this section we will first describe our simulation set up,  and make some checks.

\subsection{Simulation}\label{simulation}

We carry out our cosmological simulations by the publicly available code GADGET-2 \citep{Gadget2,Gadget1}\footnote{\url{https://wwwmpa.mpa-garching.mpg.de/gadget/}},
which uses the smoothed particle hydrodynamics (SPH) method to solve the gas dynamics equations.
The publicly available version does not involve radiative heating/cooling and the chemistries that are necessary for 
correctly modeling the gas temperature evolution. We add the evolution of the free electrons, H and He ions, and
 the associated cooling/heating processes. Free electrons are essential in the global IGM temperature evolution. 
The initial free electron abundance and gas temperature are computed from the epoch of recombination using  the RECFAST code
\citep{Seager1999}\footnote{\url{https://www.astro.ubc.ca/people/scott/recfast.html}}.
We ignore here the formation of H$_2$  and the cooling it induced, as we are focusing on the nonlinear structures that have not yet 
experienced star formation processes. The homogeneous gas temperature and ionization state is evolved by solving the equations
\begin{align}\label{Eq.homoT}
\frac{{\rm d}T_{\rm K}}{dt} &= -2H(z)T_{\rm K}-\frac{2[\Lambda_{\rm net}(T_{\rm K})]}{3k_B n_{\rm tot}}, \nonumber \\
\frac{dn_{\rm HII} }{dt} =& -3n_{\rm HII}H(z)+\gamma_{\rm HI}(T_{\rm K}) n_{\rm HI}n_e \nonumber \\
 & -\alpha_{\rm HII}(T_{\rm K})n_{\rm HII}n_e, \nonumber \\
\frac{dn_{\rm HeII}}{dt} =& -3n_{\rm HeII}H(z)+\gamma_{\rm HeI}(T_{\rm K}) n_{\rm HeI}n_e \nonumber \\
 & -\alpha_{\rm HeII}(T_{\rm K}) n_{\rm HeII}n_e, \nonumber \\
\frac{dn_e}{dt}&=\frac{dn_{\rm HII}}{dt}+\frac{dn_{\rm HeII}}{dt},
\end{align}
where  $n_{\rm HI}$, $n_{\rm HII}$, $n_{\rm HeI}$, $n_{\rm HeII}$ and $n_e$ are the physical number densities of 
neutral hydrogen,  ionized hydrogen, neutral helium, singly ionized helium and electron, respectively, He III is neglected here. 
$\alpha_{\rm HII}$ and $\alpha_{\rm HeII}$ are the recombination rates, 
$\gamma_{\rm HI}$ and $\gamma_{\rm HeI}$ are the collisional ionization rates which are in fact negligible, and  
$\Lambda_{\rm net}$ is the net cooling rate. 
In the gas temperature evolution we include  the Compton scattering and Bremsstrahlung,  as detailed in \citet{CRASH2003}.

\begin{figure}[htb]     
\centering{
\includegraphics[scale=0.5]{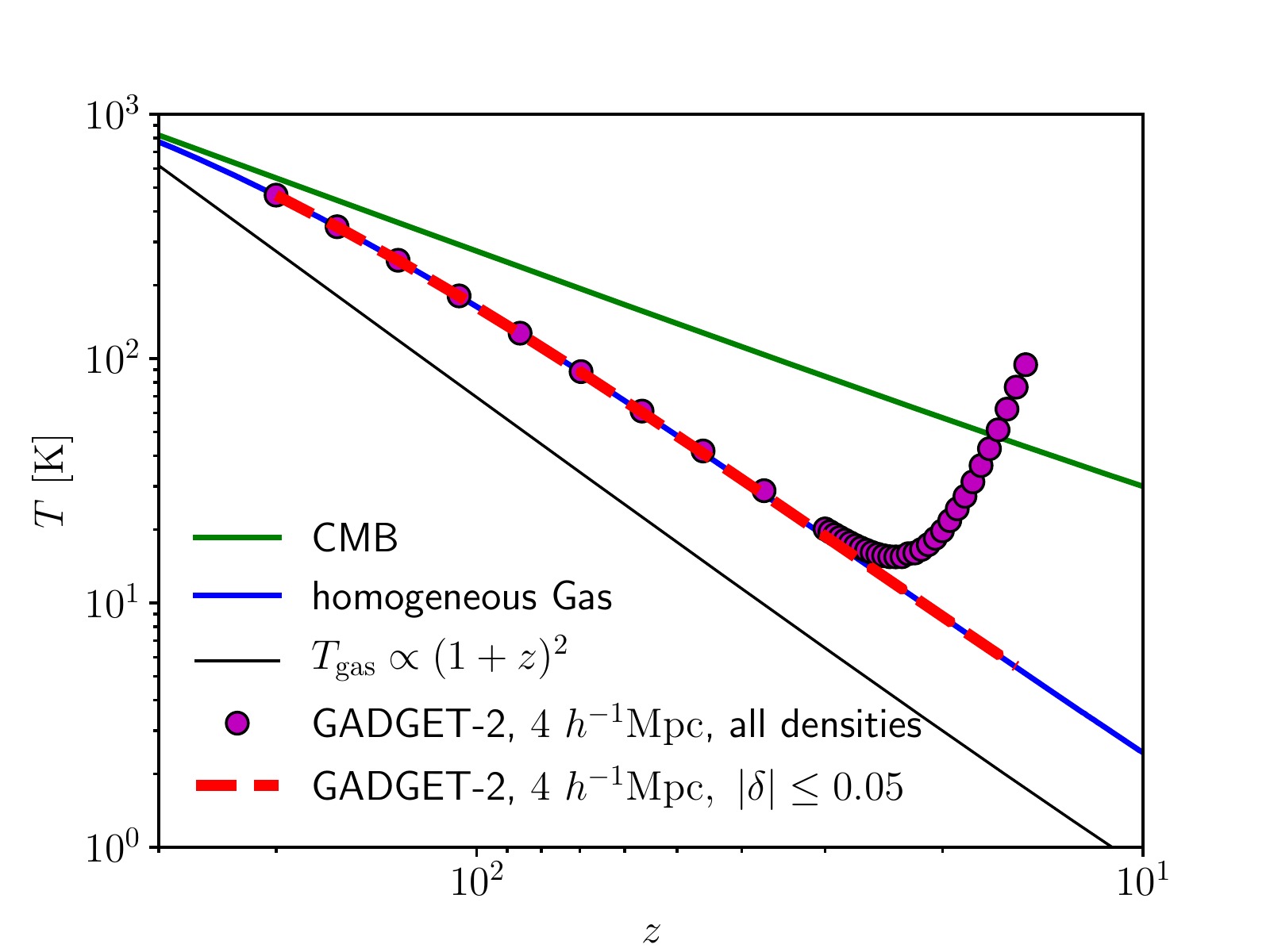}
\caption{The evolution of the mean gas temperature in our fiducial simulation. 
The filled circles show the evolution of the mean temperature of all gas particles, and the dashed line shows the average over 
gas particles with density contrast $|\delta| < 0.05$. For comparison, the evolutions of the CMB temperature and 
the homogeneous gas temperature are plotted with the green and blue solid lines respectively, and 
the thin black solid line shows a purely adiabatic evolution, shifted with an arbitrary amplitude to ease comparison.}
\label{Fig.T}
}
\end{figure}

In Fig. \ref{Fig.T} we plot the evolution of the mean gas temperature in a simulation that has a box size of 4 Mpc/$h$ and $800^3$ dark matter particles and $800^3$ gas particles respectively.    
The filled circles show the arithmetic mean temperature of all gas particles, 
including those within and near halos that are shock-heated,
while the dashed line corresponds to the average over 
gas particles with density contrast $|\delta| < 0.05$, which are likely less affected by the shock-heating.

\begin{figure*}[htb]     
\centering{
\includegraphics[scale=0.38]{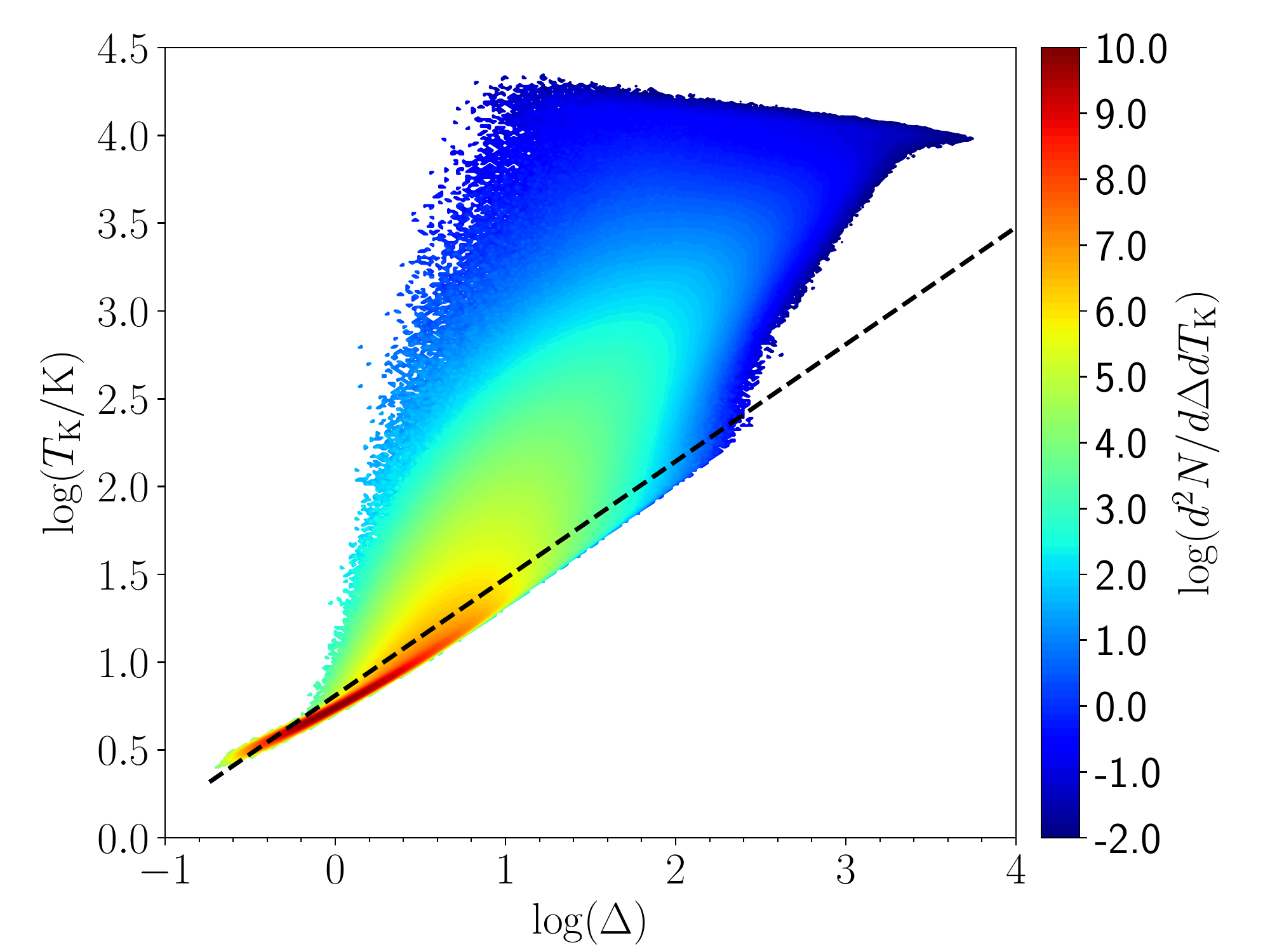}
\includegraphics[scale=0.38]{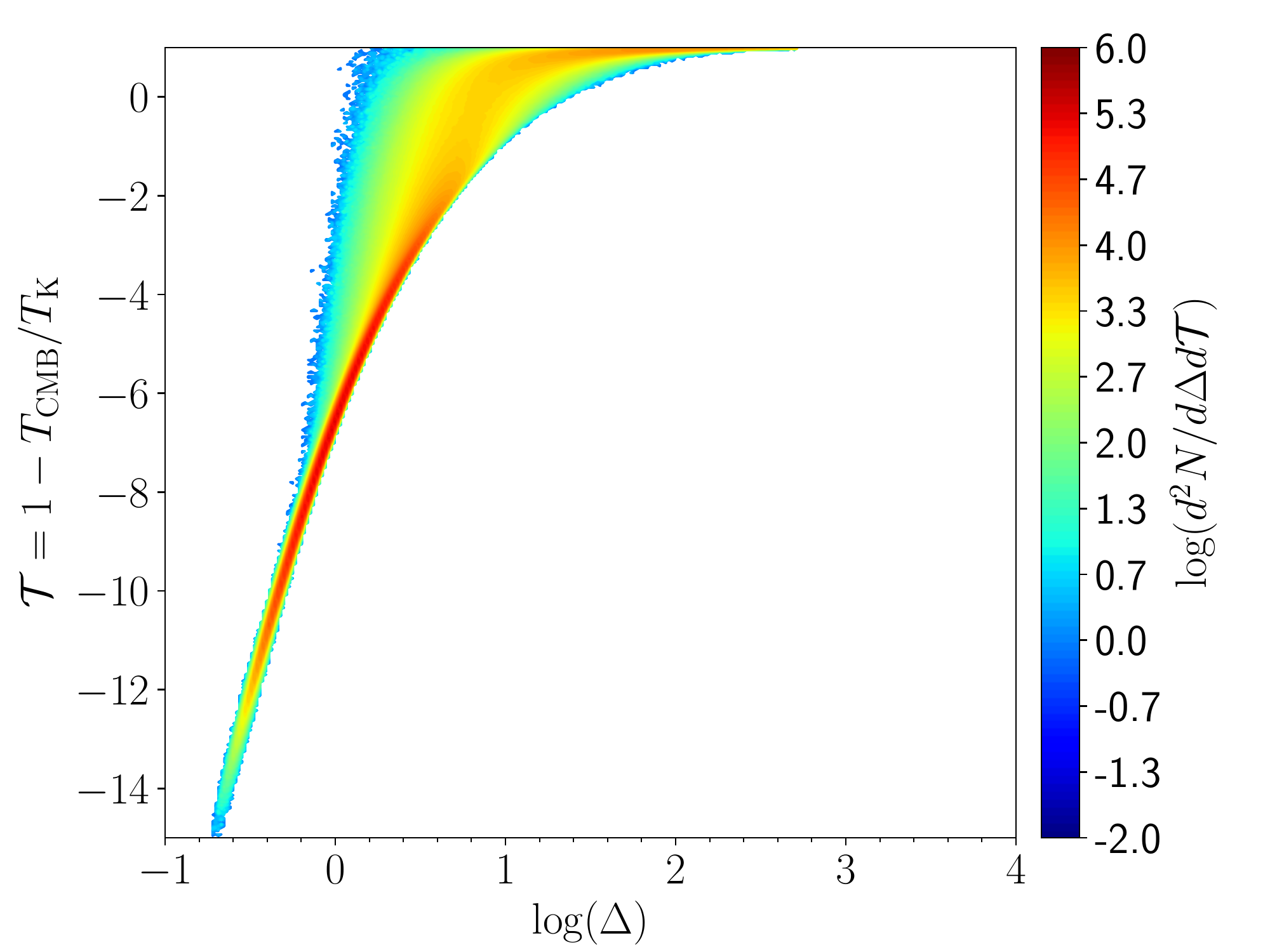}
\caption{The $T_{\rm K}-\Delta$ relation ({\it left panel}) and the $(1-T_{\rm CMB}/T_{\rm K}) - \Delta$ relation 
({\it right panel}) of gas particles at $z=17$ from our fiducial simulation. The color denotes the probability distribution of
particles at a certain position on the plane.
 As a comparison, we plot a curve of $T_{\rm K}=6.46\Delta ^{2/3}$ K with the dashed line in the left panel.}
\label{Fig.T-Delta}
}
\end{figure*}

The expected temperature evolution for homogeneous gas solution of Eq.(\ref{Eq.homoT})) is plotted with the blue solid line in the same figure.
The simulation correctly captures the global temperature evolution trend
for the mean-density gas,
and it shows that the shock-heating effect becomes significant at $z\lesssim25$. 
 Note that the arithmetic mean temperature is significantly boosted by the small number of gas particles with high temperature in halos.
As will be discussed later (section \ref{conclusions}), there is still no real consensus on the exact magnitude of the shock heating effect. Our following results are based on the GADGET-2 simulation which adopts the SPH algorithm, and predicts a reasonable evolution of the mean gas temperature.

For monoatomic gas that experiences only adiabatic compression without shock-heating, radiative heating/cooling, or
any change of chemical species, a relation between the gas temperature and the density builds:
\begin{equation}
T_{\rm K}(\Delta)=T_0 \Delta ^{2/3},
\label{eq:T}
\end{equation}
where  $T_0$ is the temperature of the mean-density gas. 
This adiabatic relation is widely adopted when estimating the IGM temperature analytically. 
However, during the cosmic evolution, even before the formation of any luminous objects, the Compton heating 
that arise from scattering with the CMB photons and the shock-heating in over-dense regions that are undergoing 
non-linear structure formation can affect  the gas temperature and break this relation.
The left panel of Fig. \ref{Fig.T-Delta} shows the probability distribution of particles on the $T_{\rm K}-\Delta$ plane from 
our simulation at $z=17$.  The right panel shows the probability distribution of particles on the 
$(1-T_{\rm CMB}/T_{\rm K})-\Delta$ plane, which is more closely related to the 21 cm brightness temperature.  The color denotes the number density of particles in the corresponding spaces.
It is seen from the left panel that the power law relation is clearly broken in the simulation. 
For $\Delta \lesssim 1$ the deviation is mainly due to the Compton-heating, while for $\Delta \gtrsim 1$ the shock-heating is the main source of heating. It is therefore essential to take  the Compton-heating and shock-heating into account.

Note that, however, the shock-heated gas only occupies a small volume fraction.
If we define the shock-heated region as those overdense regions with temperature higher than the adiabatic temperature of Eq.(\ref{eq:T}) by a fraction of 30\%, 50\%, or 100\%, then the shock-heated volume fraction is 3.5\%, 2.5\%, and 1.5\%, respectively.

\subsection{The maximum 21 cm signal}\label{21cmspect}

\begin{figure}[htb]     
\centering{
\includegraphics[scale=0.4]{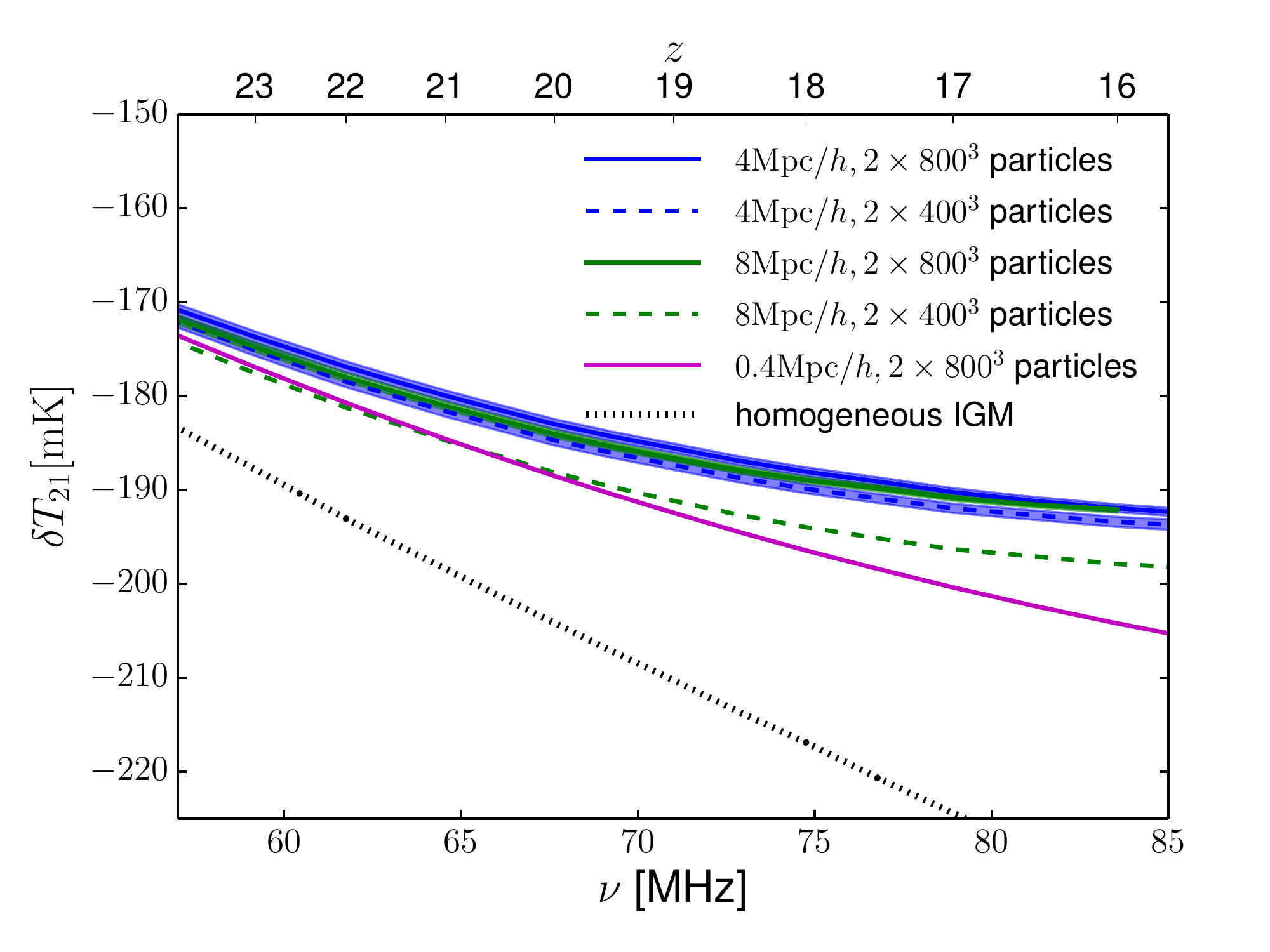}
\caption{The global 21 cm spectrum from cosmic dawn assuming saturated coupling between the
spin temperature of neutral hydrogen and the gas kinetic temperature. The blue, green, and magenta lines show results 
from simulations with box sizes of $4 \Mpc/h$, $8 \Mpc/h$, and $0.4 \Mpc/h$, respectively. The solid and dashed lines correspond
to results from simulations with particle numbers of $2\times 800^3$ and $2\times 400^3$, respectively. The black dotted line 
represents the maximum signal level expected from the homogeneous IGM. 
The shaded regions of the same color as the corresponding lines indicate the jackknife error.
}
\label{Fig.converge}
}
\end{figure}

We convert the particle field into the density field with the Cloud-in-Cell (CIC) method, and 
calculate the global 21 cm signal from the simulation.
The 21 cm brightness temperature from a uniform cell in the simulated box is
\begin{equation}
{\rm \delta}T_{\rm 21} = \frac{T_{\rm S} - T_{\rm \gamma}}{1+z}\, \left(1-e^{-\tau}\right),
\end{equation}
where $T_{\rm S}$ is the spin temperature of the neutral hydrogen, $T_{\rm \gamma}$ is the brightness temperature
of the background radiation, and $\tau$ is the 21 cm optical depth. In the absence of any extra radio background
at cosmic dawn (e.g. \citealt{2018ApJ...868...63E}), the only radio background is the CMB, 
so that $T_{\rm \gamma}(z) = T_{\rm CMB}(z)$.
As the peculiar velocity has only negligible effect on the sky-averaged 21 cm signal \citep{2018ApJ...869...42X}, 
the optical depth can be written as
\begin{equation}
\tau = \frac{3}{16}\, \frac{\hbar \,c^3\, A_{\rm 10}}{k_{\rm B}\,\nu_{\rm 21}^2}
\frac{n_{\rm HI}}{T_{\rm S}\, H(z)}.
\end{equation}
where $A_{\rm 10} = 2.85 \times 10^{-15} s^{-1}$ is the Einstein coefficient for the spontaneous decay of the 21 cm transition, 
$\nu_{\rm 21} = 1420.4 \MHz$ is the frequency of the transition, and $n_{\rm HI}$ and $H(z)$ are the local neutral hydrogen 
number density and the Hubble parameter, respectively.
In the present work, we focus on the maximum absorption signal of 21 cm that is achievable in the standard $\Lambda$CDM 
model. Therefore, in all the following calculations, we assume saturated coupling between the spin temperature of 
hydrogen and the kinetic temperature of the gas, so that $T_{\rm S} = T_{\rm K}$.
The 21 cm global signal is computed by averaging the 21 cm brightness temperature over all the cells in the simulation.

The spectra of the maximum 21 cm absorption from several
simulations are plotted in Fig.~\ref{Fig.converge}.
The shaded regions of the same color as the corresponding lines indicate the jackknife errors in the 
corresponding spectra.
The expected spectrum from the homogeneous IGM,  with the homogeneous solution for the gas temperature 
(the blue solid line in Fig.~\ref{Fig.T}), is plotted with the dotted line for comparison. 
It is seen that the non-linear structure formation affects the 21 cm absorption level obviously; 
the homogeneous assumption of the IGM would over-estimate the absorption.
The effect gets more and more significant for lower redshifts, as more
non-linear structures form.

The solid and dashed lines with different colors in Fig.~\ref{Fig.converge} show results from 
simulations of different box sizes and resolutions.
As the redshift decreases, the large-scale perturbations becomes more and more important, and a limited box
size would under-estimate the effect of non-linear structure formation because of the delayed 
structure formation. This is seen from the solid magenta line predicted by the simulation with a box size of $0.4 \Mpc/h$,
though it has the highest resolution.
It significantly over-estimates the absorption signal at $\gtrsim 70$ MHz.
At the same time, the influence of non-linear structure formation becomes more significant, and the IGM density
fluctuations become more non-linear as the redshift decreases. An insufficient resolution would 
also under-estimate the effect of non-linear structure formation by losing small structures, and this is shown by the
green dashed line in the figure from a simulation with a box size of $8 \Mpc/h$ and a particle number of 
$2\times 400^3$.
By computing the jackknife errors on the 21 cm spectrum, we find that the convergency can be achieved 
with a simulation with a box size larger than $4 \Mpc/h$ and a particle number larger than $2\times 800^3$.
The systematic error is within 1\%. 

In the following analysis, we will take the simulation with $4\Mpc/h$ size and $2\times800^3$ particles as the
fiducial simulation. From the fiducial simulation, at $z = 17$, where the EDGES absorption trough locates, 
the 21 cm absorption signal is $-190\mK$.  
The absorption amplitude is reduced by $15 \%$ w.r.t. the homogeneous IGM case  
at this redshift, when the non-linear structure formation is taken into account.
The effects are more significant as the IGM becomes more non-linear.

\subsection{Under-resolved signal}\label{spect4semi}

In order to survey a large parameter space and investigate the various effects on the global 21 cm spectrum,
a set of semi-numerical simulations are usually used to compute the signal (e.g. \citealt{2017MNRAS.472.1915C}). 
This kind of simulations usually cover a sufficiently large volume while not having a high enough resolution
to resolve non-linear structures, such as halos and their ambient gas, though the shock heating effects could
be implemented with a sub-grid algorithm. Therefore, it is necessary to see the effect of losing small-scale structures 
while keeping large-scale fluctuations just as a semi-numerical simulation does.

\begin{figure}[htb]
\centering{
\includegraphics[scale=0.4]{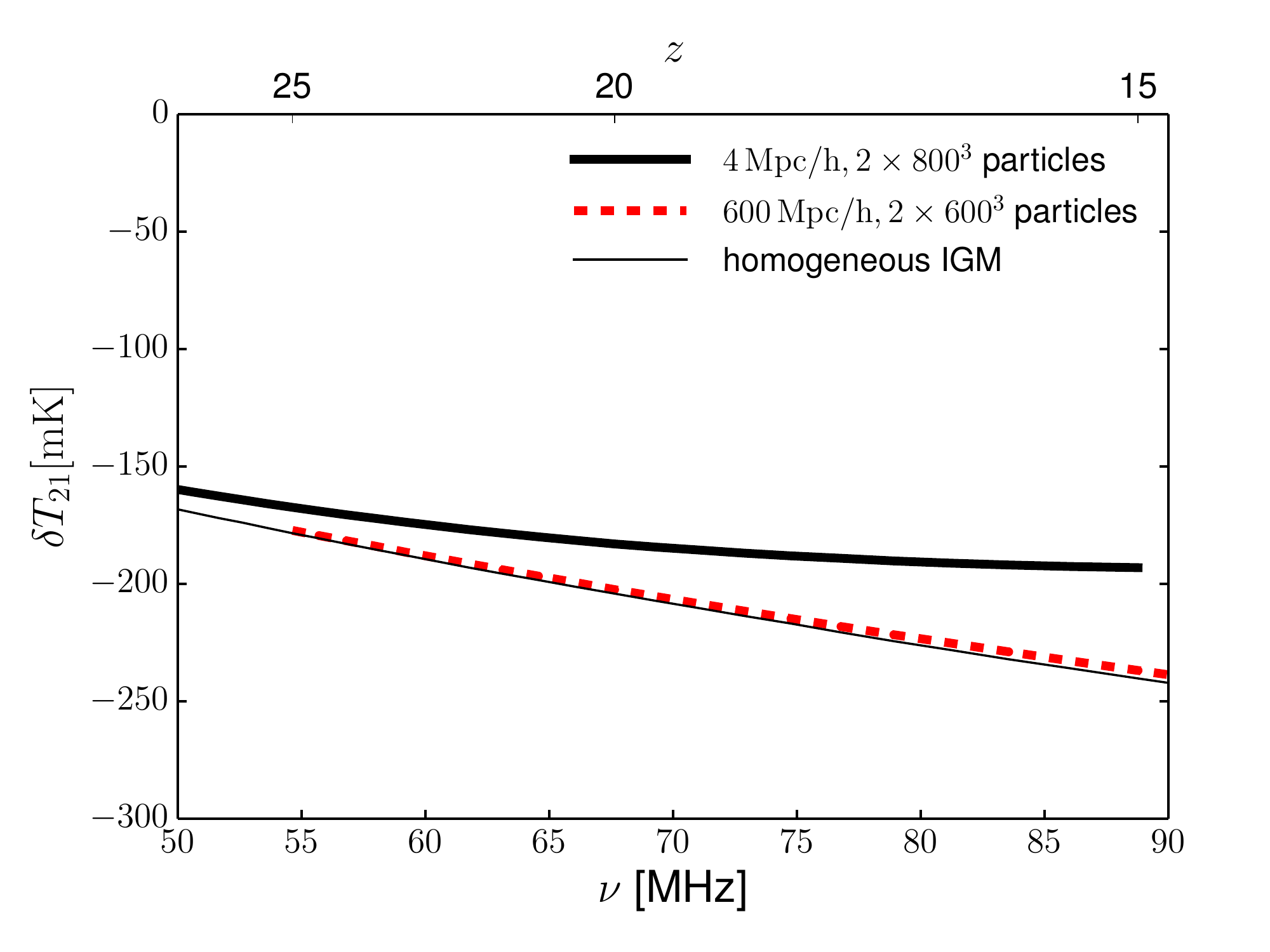}
\caption{The maximum absorption signal of the 21 cm global spectrum from cosmic dawn. The thick solid line shows the 
spectrum from our fiducial simulation, the thick dashed line shows the results from a low-resolution simulation, which has a
box size of $600 \Mpc/h$ and $2\times600^3$ particles, and the thin solid line is the maximum expectation for the 
homogeneous IGM.
}
\label{Fig.low_resol}
}
\end{figure}

\begin{figure*}[tb]     
\centering{
\includegraphics[scale=0.38]{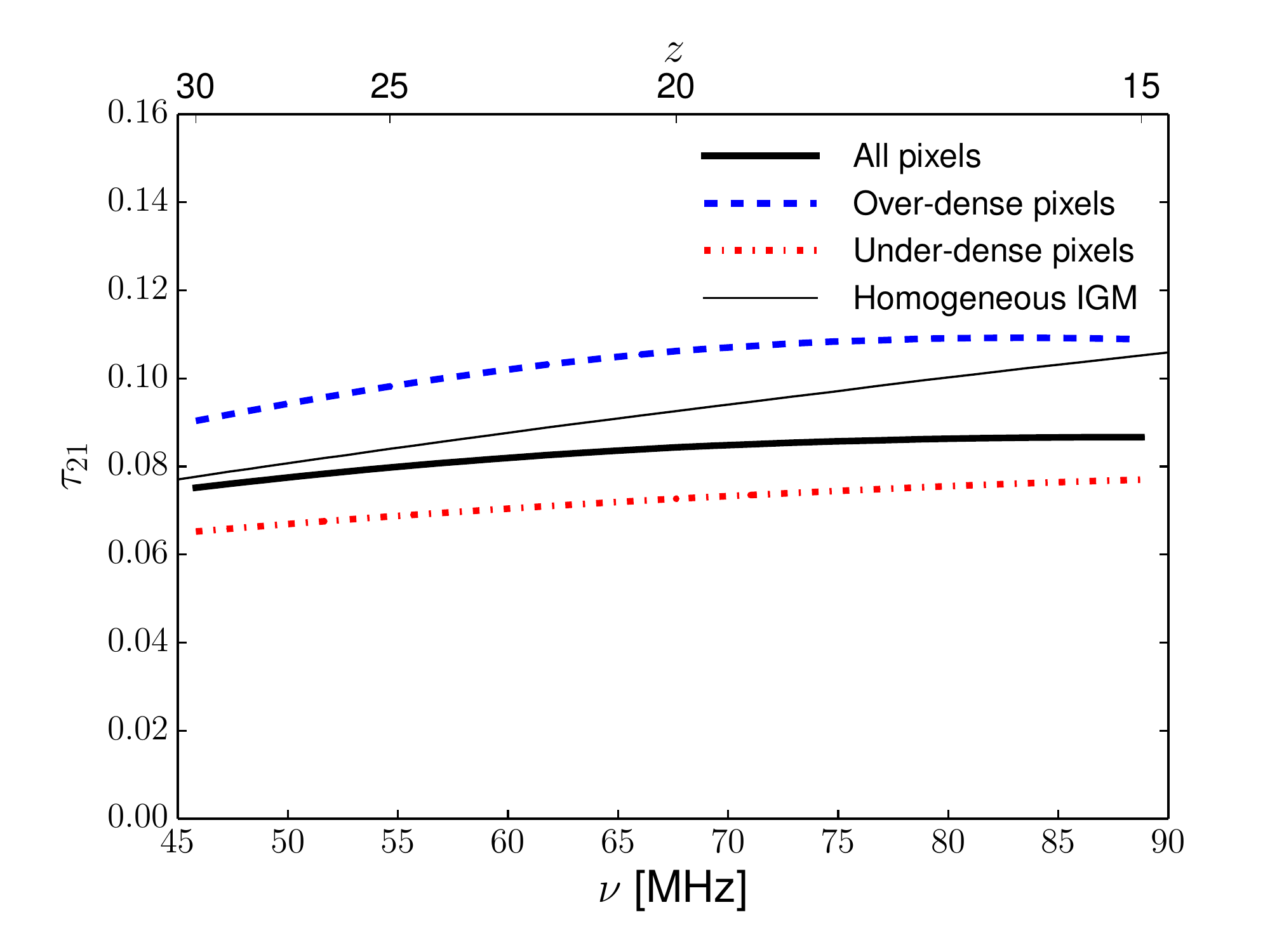}
\includegraphics[scale=0.38]{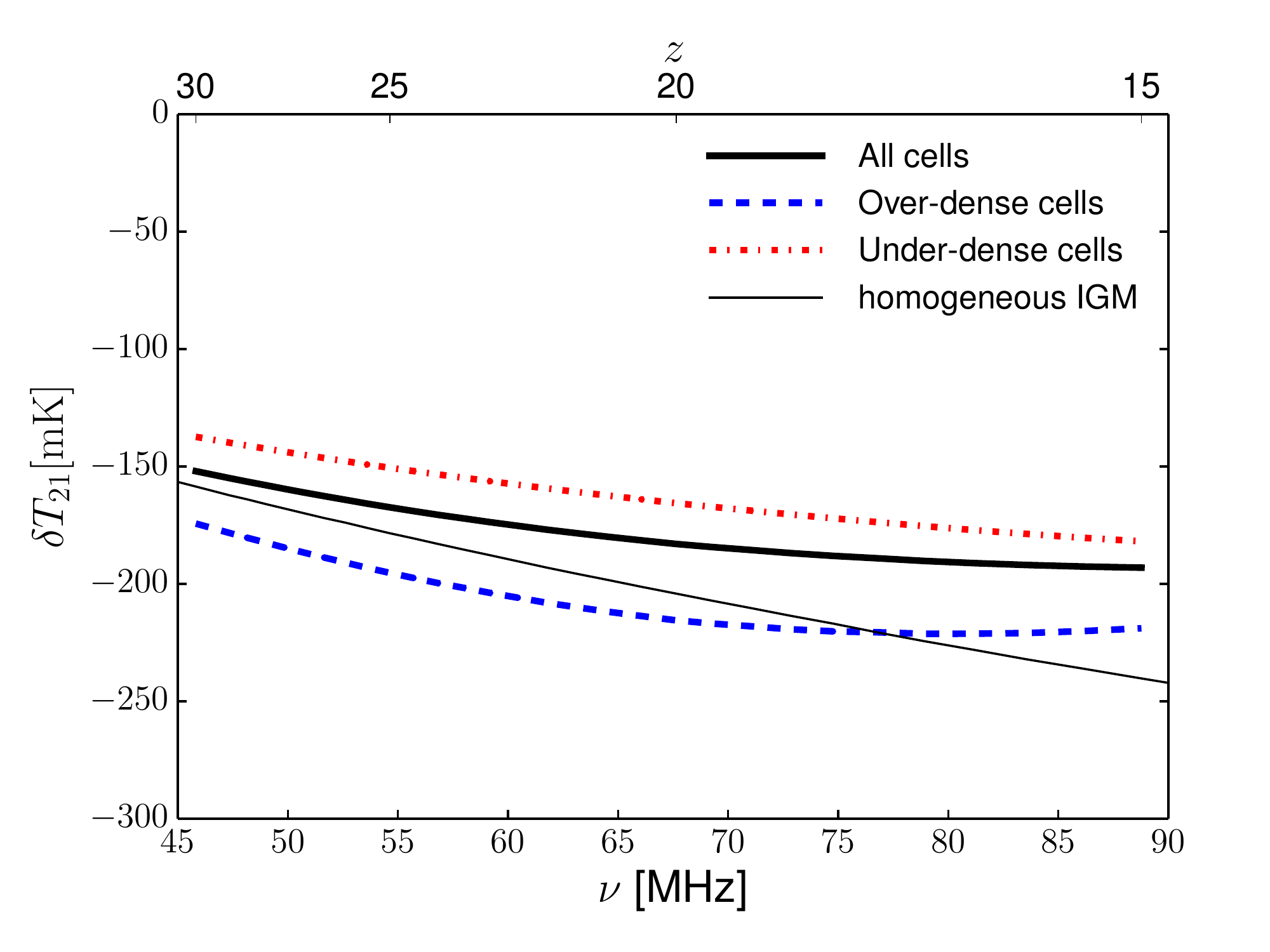}
\caption{The 21 cm optical depth ({\it left panel}) and the 21 cm global spectrum ({\it right panel}) 
averaged over over-dense pixels (all pixels with $\delta > 0$, blue dashed line) 
and that averaged over under-dense pixels (all pixels with $\delta > 0$, red dot-dashed line).
The thick solid line shows the averaged values over all pixels in the simulation, and the thin solid line in each panel
represents the spectrum expected from the homogeneous IGM.}
\label{Fig.tau_T21_overVSunder}
}
\end{figure*}

In Fig.~\ref{Fig.low_resol} we compare our high resolution hydrodynamic simulation with the low resolution ones typically used for
semi-numerical simulation. The thick dashed line shows the expected signal from a  simulation with 
a box size of $600 \Mpc/h$ and a particle number of $2\times600^3$, a typical resolution of a semi-numerical
simulation. It shows that in the low-resolution simulation with only linear density perturbations would over-estimate
the global 21 cm signal significantly, predicting an absorption level similar to the homogeneous IGM case.
Therefore, one needs to achieve a resolution of non-linear scales, or to implement a sub-grid algorithm
for the shock effects (e.g. \citealt{2004ApJ...611..642F}), to account for the small-scale effects 
on the sky-averaged signal.

\section{Dependence on different effects}
\label{effects}
We now look more closely at the various aspects, including the density and temperature dependence, and 
the large-scale clustering, that would have impacts on the maximum signal level of the global 21 cm spectrum.

\subsection{The IGM density}\label{density}

In \citet{2018ApJ...869...42X}, by assuming the analytical infall model,  
we found that the weakly nonlinear gas around collapsed halos is adiabatically heated and 
could affect the global absorption signal. Now we study the dependence on the local over-density and 
density profiles of the IGM in this section.

Fig.~\ref{Fig.tau_T21_overVSunder} shows the averaged optical depth (left panel) and the 21 cm brightness 
temperature (right panel) of all over-dense pixels with $\delta > 0$ (dashed lines) and 
those of under-dense pixels with $\delta < 0$ (dot-dashed lines), 
respectively. The averaged values over all pixels are plotted with the thick solid line.
Although the gas in the over-dense regions has a higher temperature, most still has a lower temperature than the CMB, so
they still contribute to the total absorption signal. As the dense regions have larger optical depth, they actually contribute 
more absorption signal than the gas in under-dense regions.
However, due mainly to the adiabatic heating and the shock heating during the structure formation 
(the large-scale clustering also has a minor effect, as discussed in section~\ref{clustering}), 
as more non-linear structures form at lower redshifts, the averaged absorption signal is
weaker than the expected signal for the homogeneous IGM, even if one considers only the over-dense regions,
as can be seen at the right end of the right panel of Fig.~\ref{Fig.tau_T21_overVSunder}.

The gas in halos is shock-heated to a temperature close to the halo virial temperature, suppressing
substantially its contribution to the 21 cm absorption, and the main contribution to the 21 cm absorption
signal during the cosmic dawn comes from the gas in the less-heated IGM. The formation of dark matter halos,
however, enhances the gas density surrounding them, resulting in non-linear density fluctuations in 
the vast IGM.
Here we investigate how the detailed density profiles affect the predicted global 21 cm signal.

\begin{figure*}[htb] 
\centering{
\includegraphics[scale=0.38]{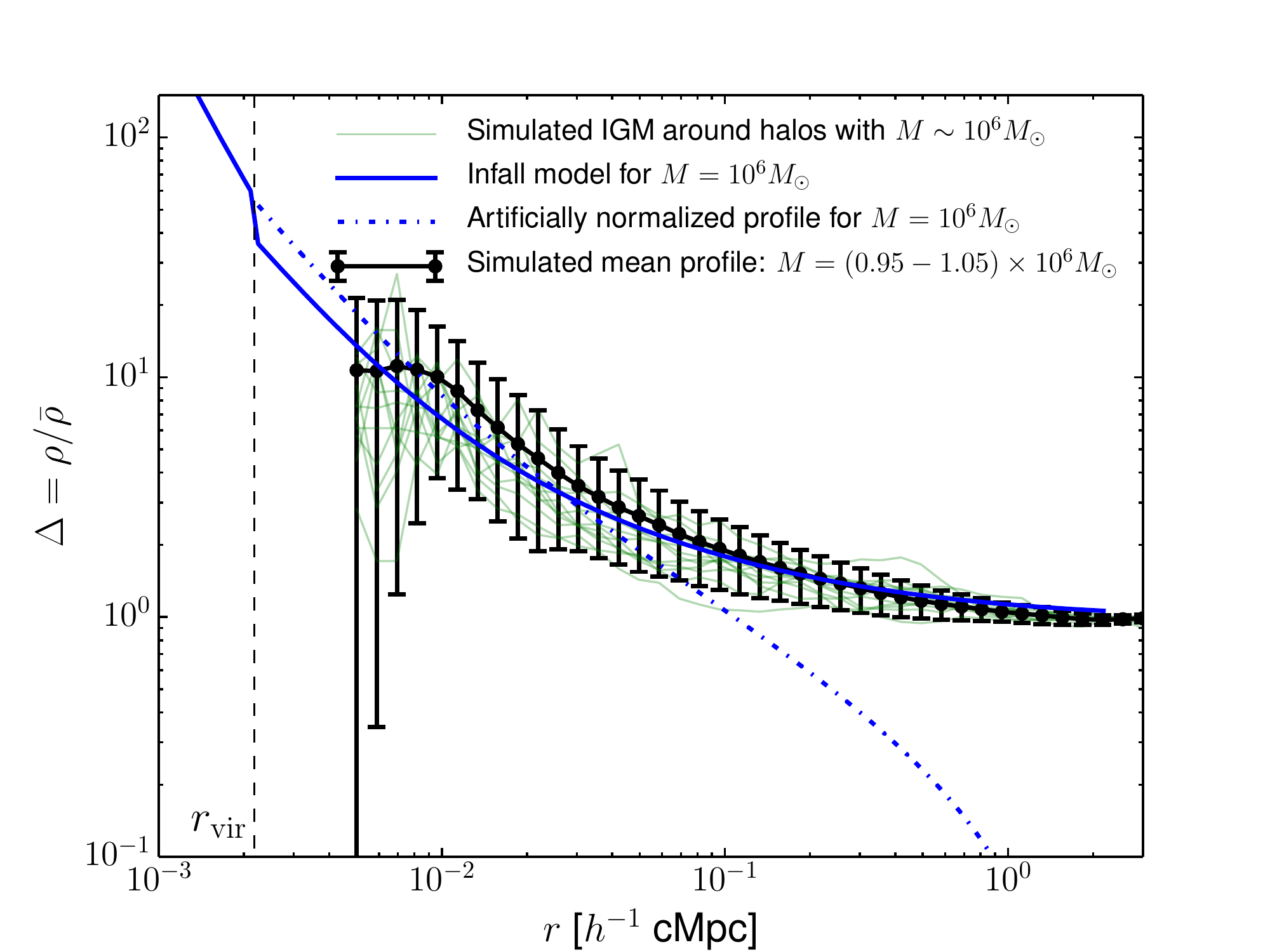}
\includegraphics[scale=0.38]{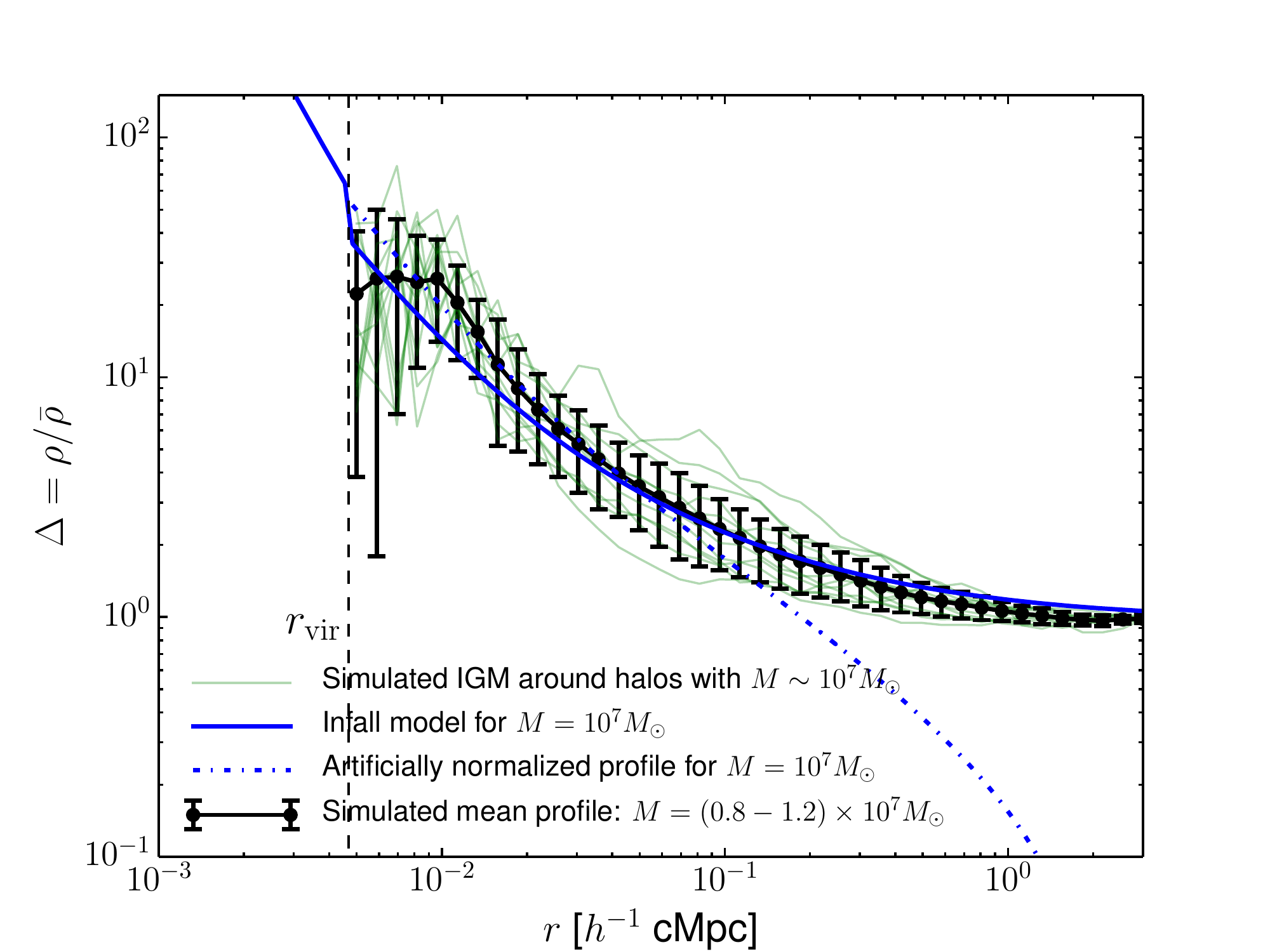}
\caption{{\it Left:} the density profiles of halos with $M \sim 10^6 M_\odot$ at redshift $z = 17$. 
{\it Right:} the density profiles of halos with $M \sim 10^7 M_\odot$ at $z = 17$.
}
\label{Fig.den_prof}
}
\end{figure*}

In Fig.~\ref{Fig.den_prof}, we plot the density profiles of the IGM  surrounding some of the individual halos 
from the hydro-dynamic simulation with light green lines, at redshift 17. 
The mean density profiles over all the halos in the mass range $(0.95 - 1.05)\times 10^6 \Msun$ (left panel), 
 and $(0.8 - 1.2)\times 10^7 \Msun$ (right panel), are plotted with black solid lines with the error bars being the standard deviations. 
 For comparison, the blue solid lines show the profiles with the corresponding halo mass and redshift predicted by the infall model, 
 which uses the excursion set theory \citep{1991ApJ...379..440B,1993MNRAS.262..627L} 
to predict analytically the density profile around 
a halo located at a density peak \citep{2004MNRAS.347...59B}.
 Generally, the density profile from the simulation is consistent with the  
 infall model, though there are large scatters in individual halos.

To assess the effect of adopting an inaccurate density profile, we populate a mock simulation box 
with the halo catalog, including the halo position and mass information, from the hydrodynamic simulation, 
but assign an artificial density profile to the IGM around each halo as prescribed by the infall model, further normalized 
such that the minimum density is zero, and the mean density equals the cosmic mean.
Note that the infall model is appropriate for halo surroundings that are located in density peaks, but may not
be applicable for under-dense environments. By introducing this artificial normalization, we mimic 
the existence of under-dense regions while keeping the cosmic mean density. 
However, we caution the readers that this artificial density profile, as shown by the blue dot-dashed lines in Fig.~\ref{Fig.den_prof},
 is unphysical, and here we only use it to investigate the effect of a steeper density profile on the global 21 cm signal.

A {\it mock} adiabatic temperature is assigned to each pixel according to the local density using
the adiabatic relation of $T_{\K} \propto \rho^{2/3}$ for the ideal gas, with the mean-density gas having a 
temperature of $6.46 \K$ at redshift 17, consistent with the homogeneous gas temperature calculated with
Eq.(\ref{Eq.homoT}) (i.e. the blue solid line in Fig.~\ref{Fig.T}).
We call this temperature the {\it mock} adiabatic temperature as it accounts for the adiabatic heating or cooling
according to the local density, but the Compton heating is taken into account when determining the mean-density
gas temperature, which is not purely adiabatic.

The resulting 21 cm signal is $-208 \mK$ ($\sim 7\%$ decrement in the absolute value 
w.r.t. the homogeneous IGM case) at $z = 17$, 
as compared to $-213\mK$ ($\sim 5\%$ decrement in the absolute value) at the same redshift if we adopt the density field
from the hydrodynamic simulation with the corresponding {\it mock} adiabatic temperature.
We find that the steeper density profile, or equivalently a higher level of density fluctuations, 
results in weaker absorption in the 21 cm signal, but the effect is only moderate.

\subsection{Gas Temperature}\label{temperature}

\begin{figure*}[htb]     
\centering{
\includegraphics[scale=0.5]{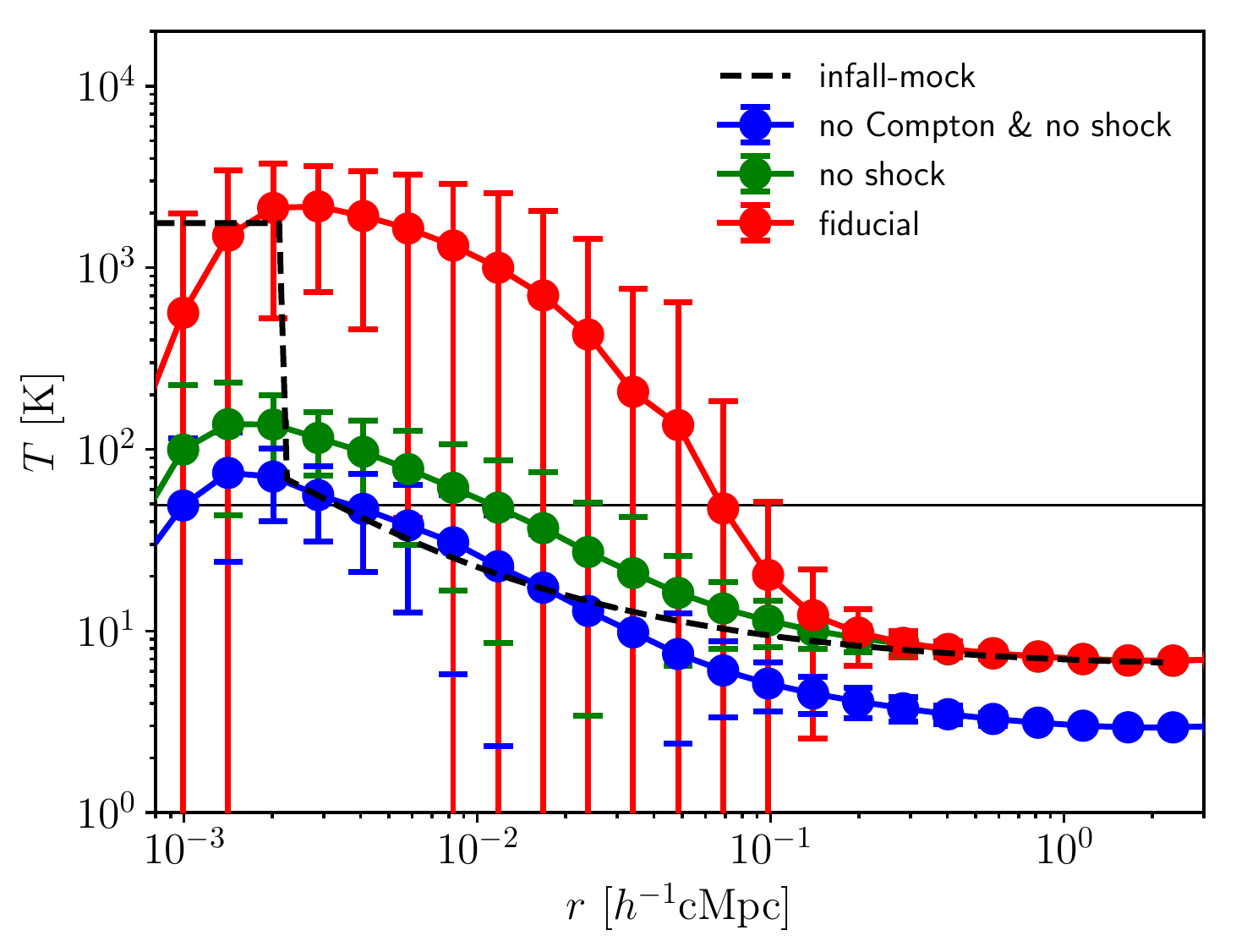}
\includegraphics[scale=0.5]{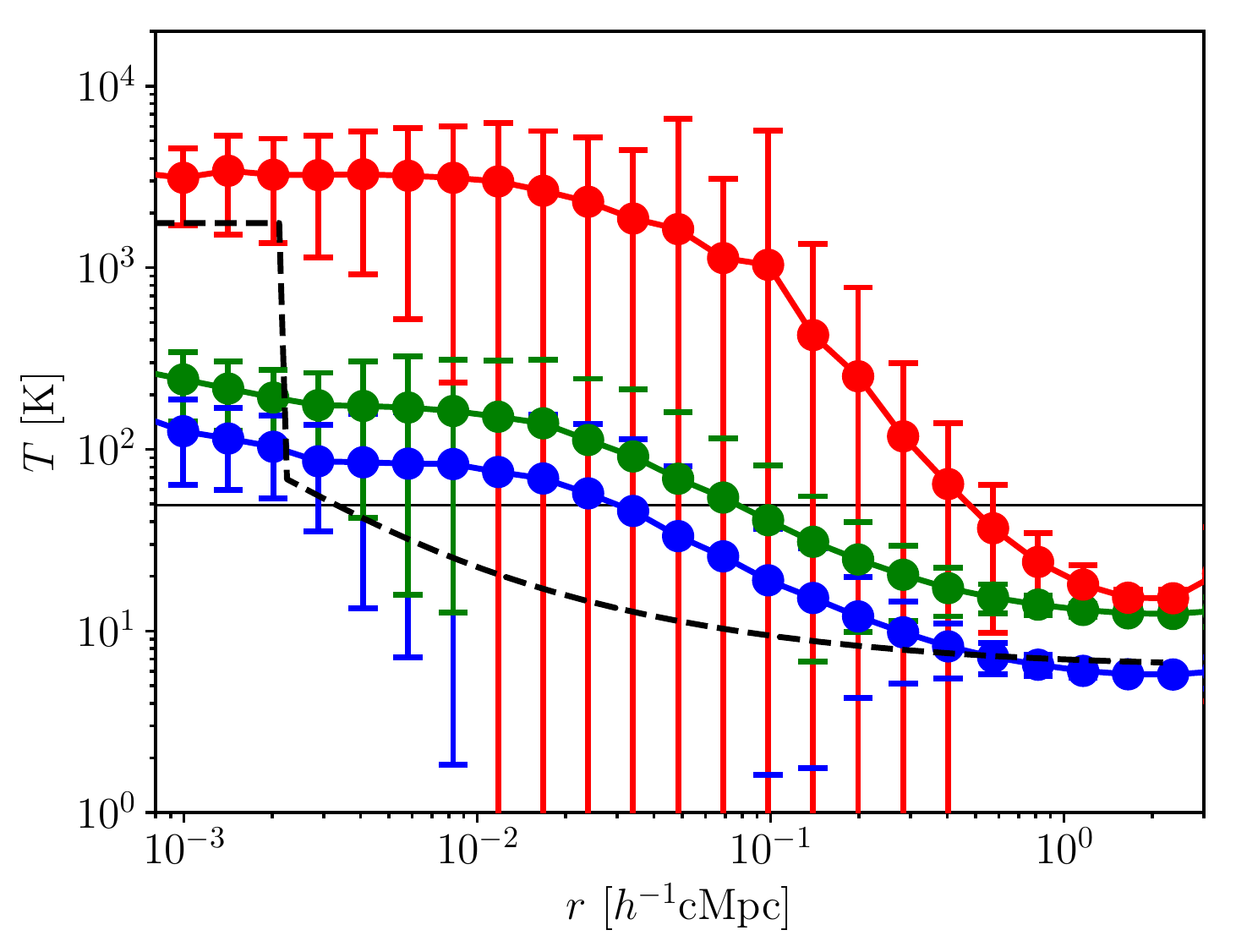}
\includegraphics[scale=0.5]{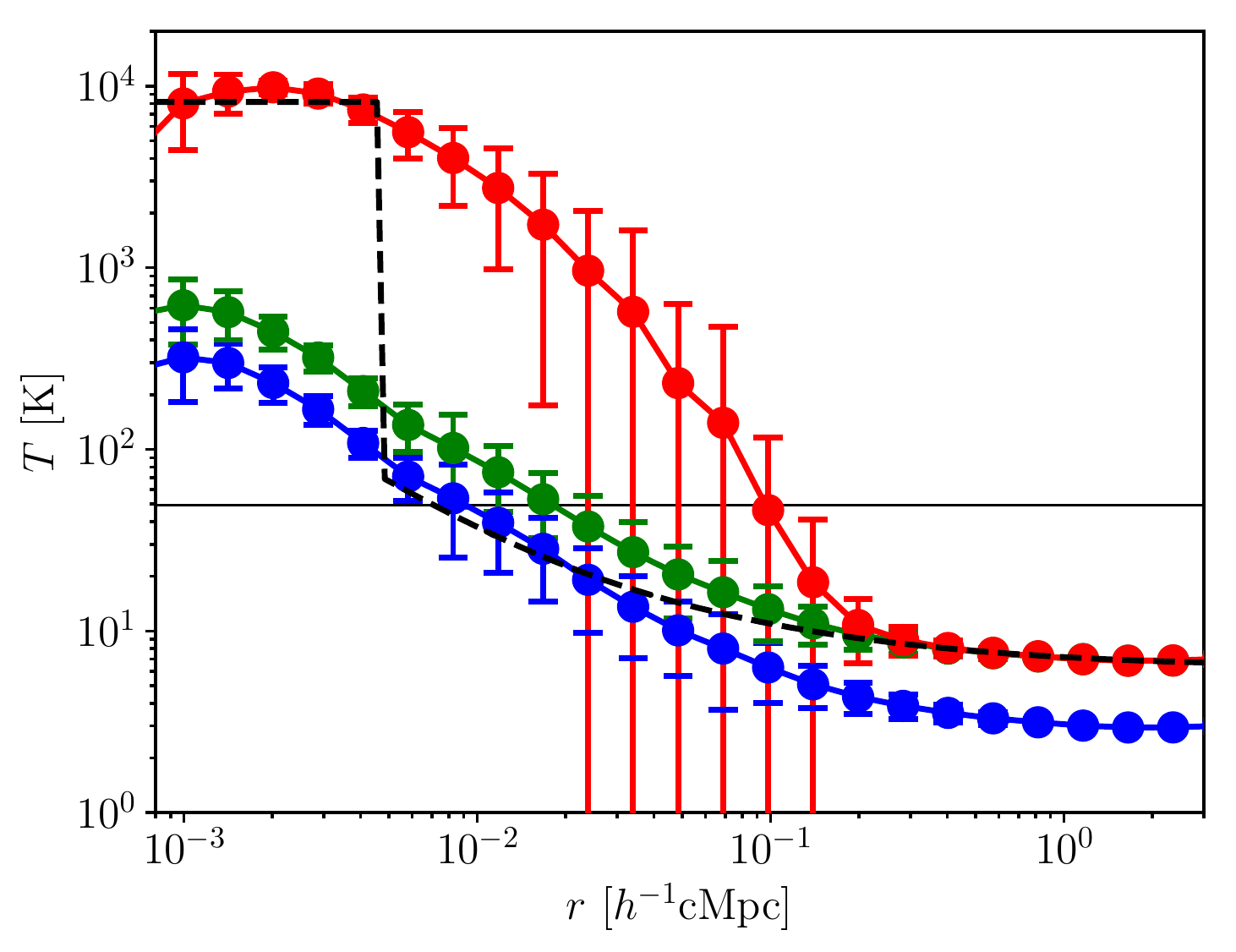}
\includegraphics[scale=0.5]{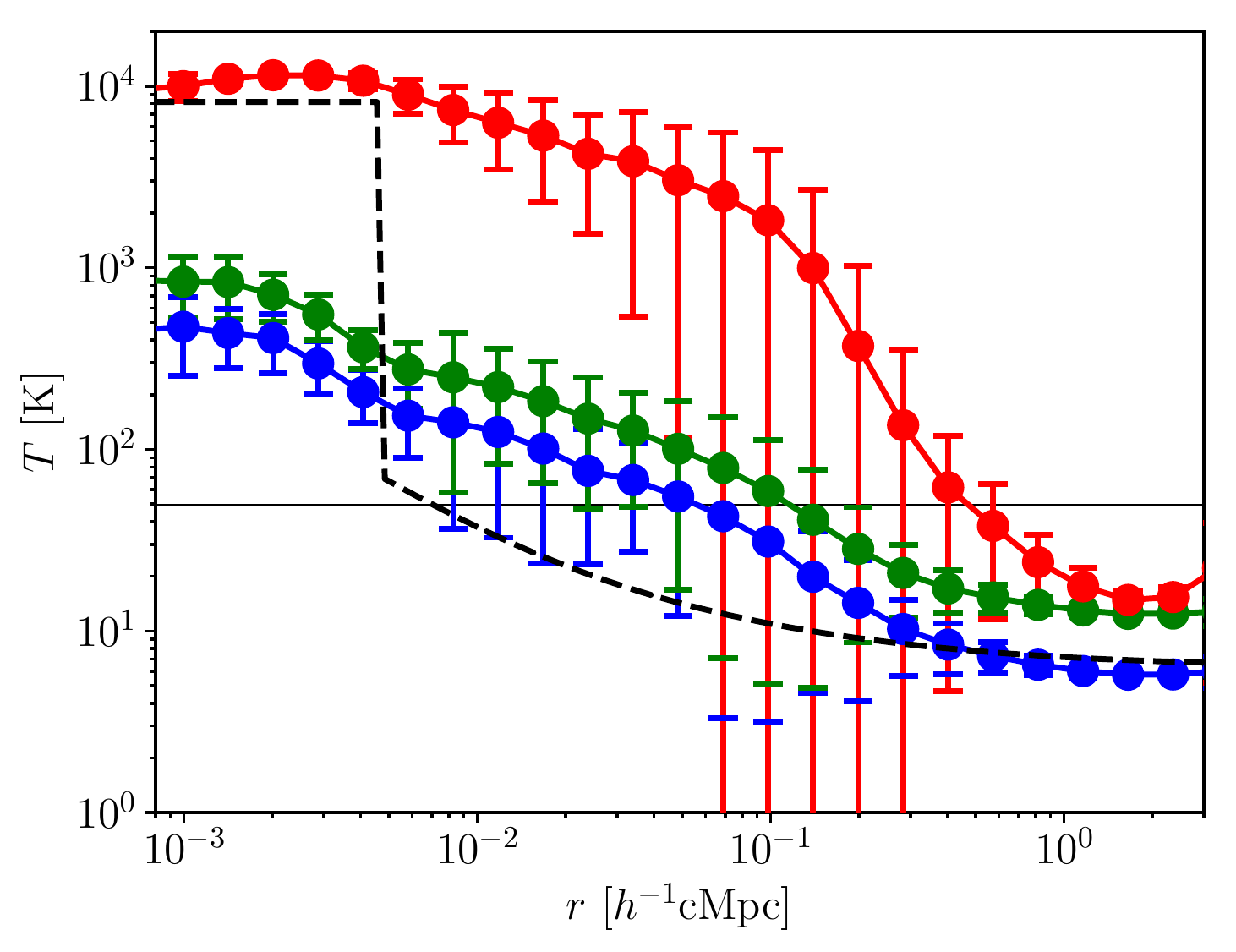}
\caption{The gas temperature profiles around a $\sim10^6~M_\odot$ halo (Top panels) and a $\sim10^7~M_\odot$ halo (Bottom panels) 
at redshift 17. In each row, the Left panel shows the profiles of the median gas temperature, and the Right panel shows the upper 
boundaries of the 90\% probability in the temperature distribution. 
The red, green, and blue curves correspond to the fiducial case, the case without 
shock-heating, and the case with both shock-heating and Compton-heating removed, respectively. 
Dashed line represents the mock-adiabatic temperature of the infall model, and the thin solid lines shows the CMB temperature.}
\label{Fig.T_prof}
}
\end{figure*}

In section~\ref{simulation}, we have seen that the shock-heating and Compton-heating can make the gas temperature 
deviate significantly from the $T_k\propto \Delta ^{2/3}$ relation (Fig.~\ref{Fig.T-Delta}). 
These effects may play a significant role in determining the 21 cm signal level during the cosmic dawn.
Here we run another two simulations with the same initial condition as the fiducial simulation; in one simulation,
we turn off the shock-heating (in GADGET-2, this is controlled by a parameter named ``artificial viscosity''), 
and in the other one, we remove both the shock-heating and the Compton heating term. 
We make a detailed comparison between the temperature profiles from the hydrodynamic simulations 
with or without these heating effects and test their impacts on the 21 cm signal.

\begin{figure}[htb]
\centering
\includegraphics[scale=0.4]{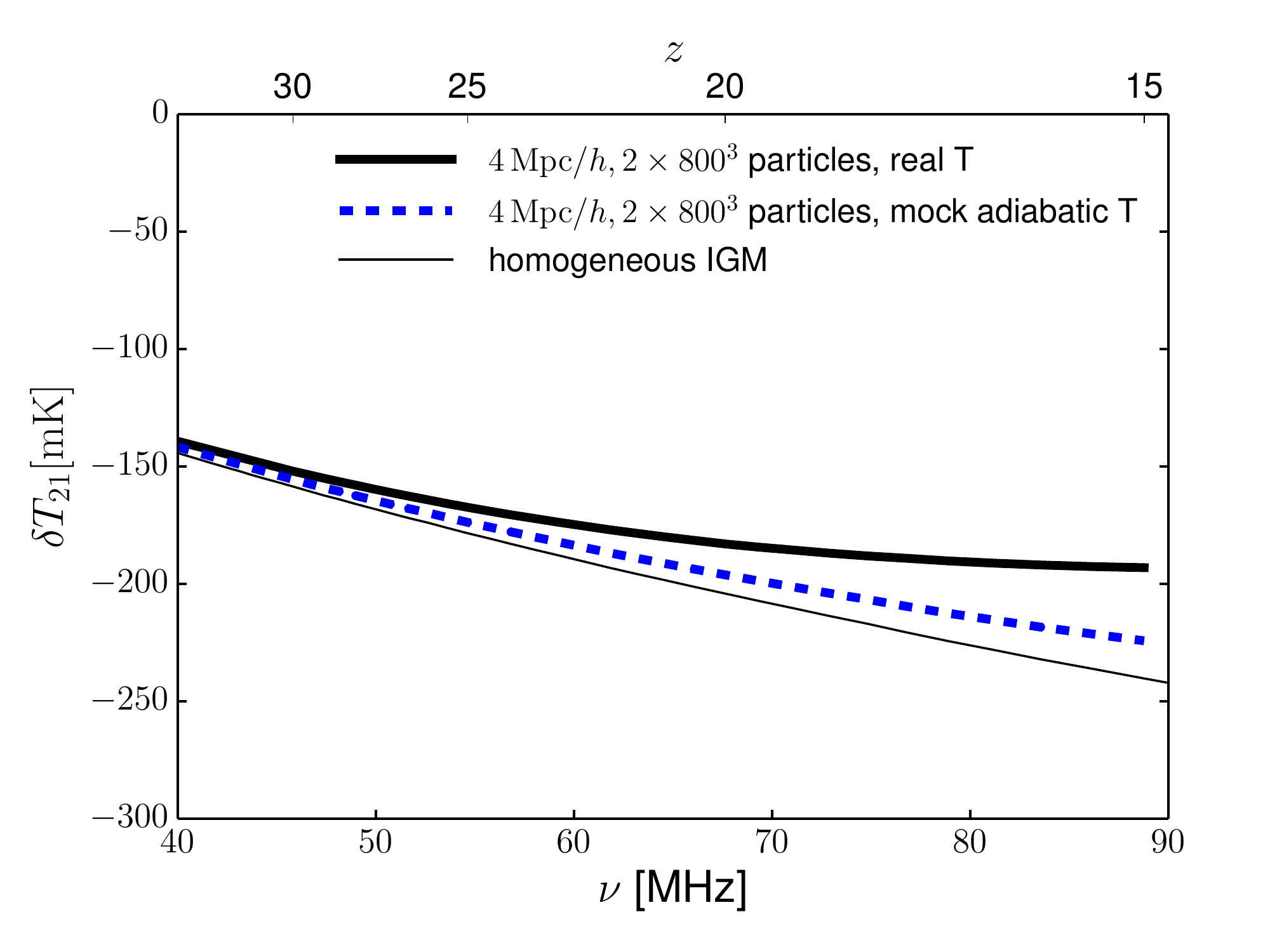}
\caption{The 21 cm global spectrum, for the simulated gas temperature ({\it thick solid} line),  mock adiabatic temperature 
({\it thick dashed} line), and the homogeneous IGM ({\it thin solid line}).}
\label{Fig.adiabaticT}
\end{figure}

Fig.~\ref{Fig.T_prof} shows the gas temperature profiles of the IGM surrounding halos with mass
$ \sim 10^6 M_\odot$ (top panels) and $ \sim 10^7 M_\odot$ (bottom panels) at redshift $z = 17$. 
For each panel we plot the temperature profiles from the fiducial simulation, the simulation without shock-heating, 
and the simulation without either shock-heating or Compton-heating, respectively. 
In each row, the left panel shows the median gas temperature at the various distances with the error bars being 
the standard deviation of the median temperature, while the right panel shows the 90\% upper limit of the temperature 
distributions and the corresponding standard deviation. 
As a reference, we also plot the {\it mock} adiabatic temperature profile derived from the infall model. 
Although the scatter among halos is quite large, it is clear that the shock heating significantly increases the gas 
temperature in over-dense regions near halos, and
the Compton heating dominates the heating effect in the vast under-dense regions.
As a result, the adiabatic assumption for the IGM temperature would
substantially over-predict the 21 cm absorption level, as shown in 
Fig.~\ref{Fig.adiabaticT} which compares the global 21 cm spectrum from the fiducial simulation (thick solid line) 
with the mock adiabatic gas temperature (dashed line).
Because of the shock heating and the Compton heating incorporated in the hydro-dynamic simulation,
the 21 cm signal at redshift 17 is further suppressed to be $-190\mK$ ($\sim15\%$ decrement in the absolute value 
w.r.t. the homogeneous IGM case), as compared to $-213\mK$ ($\sim 5\%$ decrement in the absolute value) 
in the case considering only
the gas density fluctuations and the mock adiabatic temperature.

Interestingly, in Fig.~\ref{Fig.T_prof} we find that between $\sim0.01-1~h^{-1}\Mpc$, the temperature profile 
in the fiducial simulation has large scatters. It implies that the gas in these regions experienced complicated dynamical processes. 
The shock heating makes this over-dense gas optically thin for the 21 cm signal.
However, it is very hard to observationally resolve these shocked regions. 
During the cosmic dawn the shock-heated regions only occupy a small volume fraction, but they result in 
$\sim 10\%$ decrement in the global 21 cm signal. 
The global 21 cm signal may provide some clue on these shock effects.

\begin{figure}[htb]     
\centering{
\includegraphics[scale=0.38]{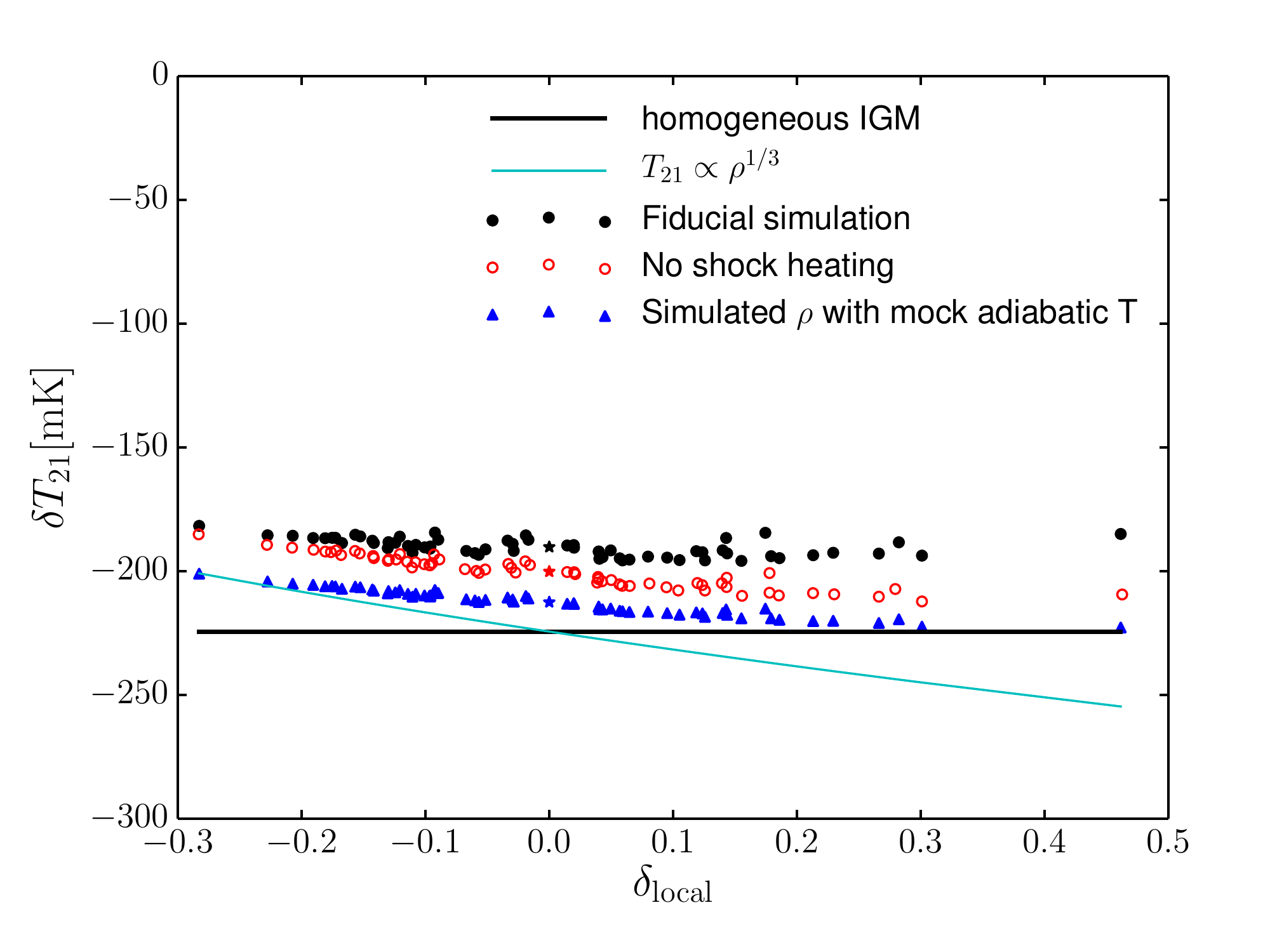}
\caption{The mean 21 cm brightness temperature of subboxes with different mean local densities 
at redshift $z = 17$.
The black dots are the $\delta T_{\rm 21}$ of subboxes from the fiducial simulation, 
the red circles are from the simulation with no shock heating, and the blue
triangles are the $\delta T_{\rm 21}$ of the subboxes assuming densities from the simulation 
but with mock adiabatic temperatures. 
The black line indicates the $\delta T_{\rm 21}$ expected from the homogeneous IGM, and the cyan line
represents the $\delta T_{\rm 21} \propto \rho^{1/3}$ scaling.}
\label{Fig.T21_subbox}
}
\end{figure}

To investigate the heating effect 
in regions with different densities, we divide the box into 
$4\times4\times4$ subboxes. Each subbox has the same size of $1\Mpc/h$ and a different local density $\delta_{\rm local}$. 
In Fig.~\ref{Fig.T21_subbox} we plot the mean 21 cm brightness temperature of the subboxes as a function of their local 
mean overdensity at $z=17$, for our fiducial simulation (black dots), 
and for the case with the fiducial density field and the mock adiabatic temperature (blue triangles).
The stars of the corresponding colors indicate the average values over the whole box.
The black solid line in the figure indicates the 21 cm brightness expected
from a homogeneous IGM for comparison, while
the cyan curve shows the 21 cm signals with the scaling of $\delta T_{\rm 21} \propto \rho^{1/3}$, 
which is expected for the cold gas ($T_{\rm S} \ll T_{\rm CMB}$) in the linear regime.
We find that at under-dense regions, the scaling between the 21 cm brightness and the local overdensity is close to 
$\delta T \propto \rho^{1/3}$ if we assume the mock adiabatic temperature, while at over-dense regions, the 
relation deviates from this scaling significantly. 
The Compton-heating and shock-heating effects further suppress the 21 cm signal, which is more prominent
in over-dense regions.

For the fiducial simulation and the mock adiabatic case, we find that a relation between the maximum 21 cm signal 
and the local overdensity of the form
\begin{equation}
\delta T_{21}= - \alpha\Delta_{\rm local}^\beta \mK
\end{equation}
holds between $-0.2<\delta_{\rm local}<0.2$. 
At redshift 17, we find $\log_{\rm 10}\alpha=2.28$ and $\beta=0.091$ for the fiducial simulation, 
while for the mock adiabatic case, we have $\log_{\rm 10}\alpha=2.33$ and $\beta=0.165$.
The heating effects result in a weaker dependence of the maximal signal on the local overdensity. 
We also fit a redshift dependence of the parameters, i.e. $\log_{\rm 10}\alpha = 2.66 -0.31\,\log_{\rm 10}(1+z)$, and
$\beta = -1.540+1.299\,\log_{\rm 10}(1+z)$, for redshifts from 25 to 15.
This fitted relation could be used in semi-numerical simulations with large box sizes but low resolutions, which would 
not be able to capture the shock-heating and non-linear density fluctuations.
Note, however, that this fitted relation applies only to the $1\Mpc/h$ cells.
A different smoothing scale would have a different scaling, and that would require a separate simulation with
a relevant box size and resolution.

\subsection{Shock-heating and Compton-heating}\label{shock}

\begin{figure}[htb]
\centering{
\includegraphics[scale=0.4]{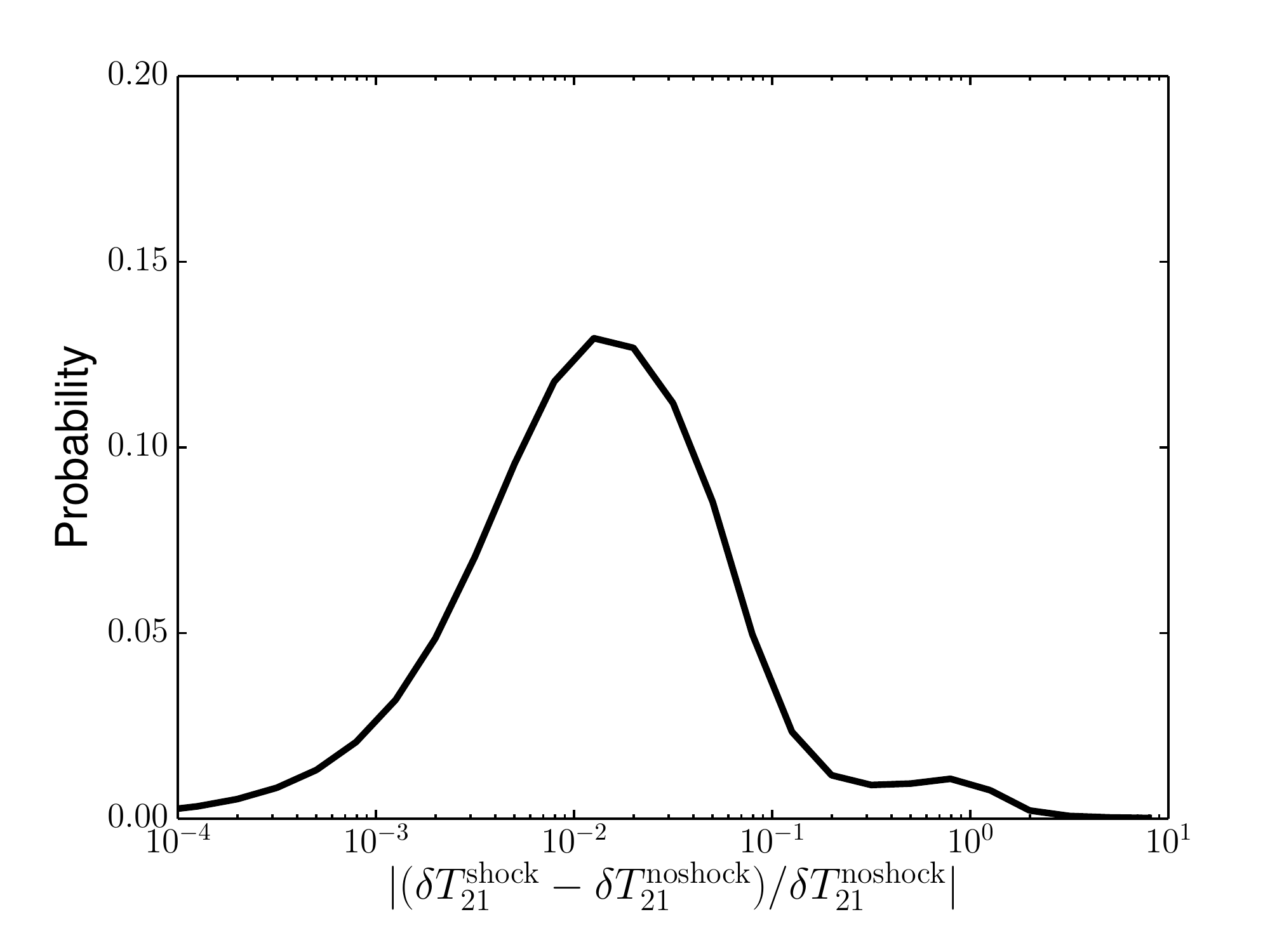}
\caption{The probability distribution of the fractional difference in the 21 cm brightness temperature between
the simulations with and without shock heating.
}
\label{Fig.shockPDF}
}
\end{figure}

To distinguish the effect of shock-heating and Compton-heating, we also compare with a simulation in which only the 
shock-heating is turned off, 
the results  at $z=17$ are plotted as the red circle symbol in Fig.~\ref{Fig.T21_subbox}.
The mean 21 cm brightness temperature averaged over the whole simulation box
is $-200\mK$, which is about 10\% decrement  in the absolute value w.r.t. the homogeneous IGM case.
Comparing the various cases, we see the shock-heating and the Compton-heating have comparable 
effects in decreasing the 21 cm absorption signal, and this is consistent with the previous study by 
\citet{McQuinn2012}. In the absence of radiation sources, the shock-heating dominates the heating effects 
in over-dense regions, while the Compton-heating dominates the heating effect in under-dense regions.

Fig.~\ref{Fig.shockPDF} shows the probability distribution of the fractional difference of the 21 cm brightness temperature
between the pixels in the default simulation and the corresponding pixels in the simulation without shock-heating.
For most pixels, the shock heating results in a suppression of the 21 cm signal by a few percent, but there is a small 
fraction of pixels that are shock-heated significantly, resulting in a long tail in the 
probability distribution.

\subsection{Halo Clustering}\label{clustering}

\begin{figure*}[htb]     
\centering{
\includegraphics[scale=0.38]{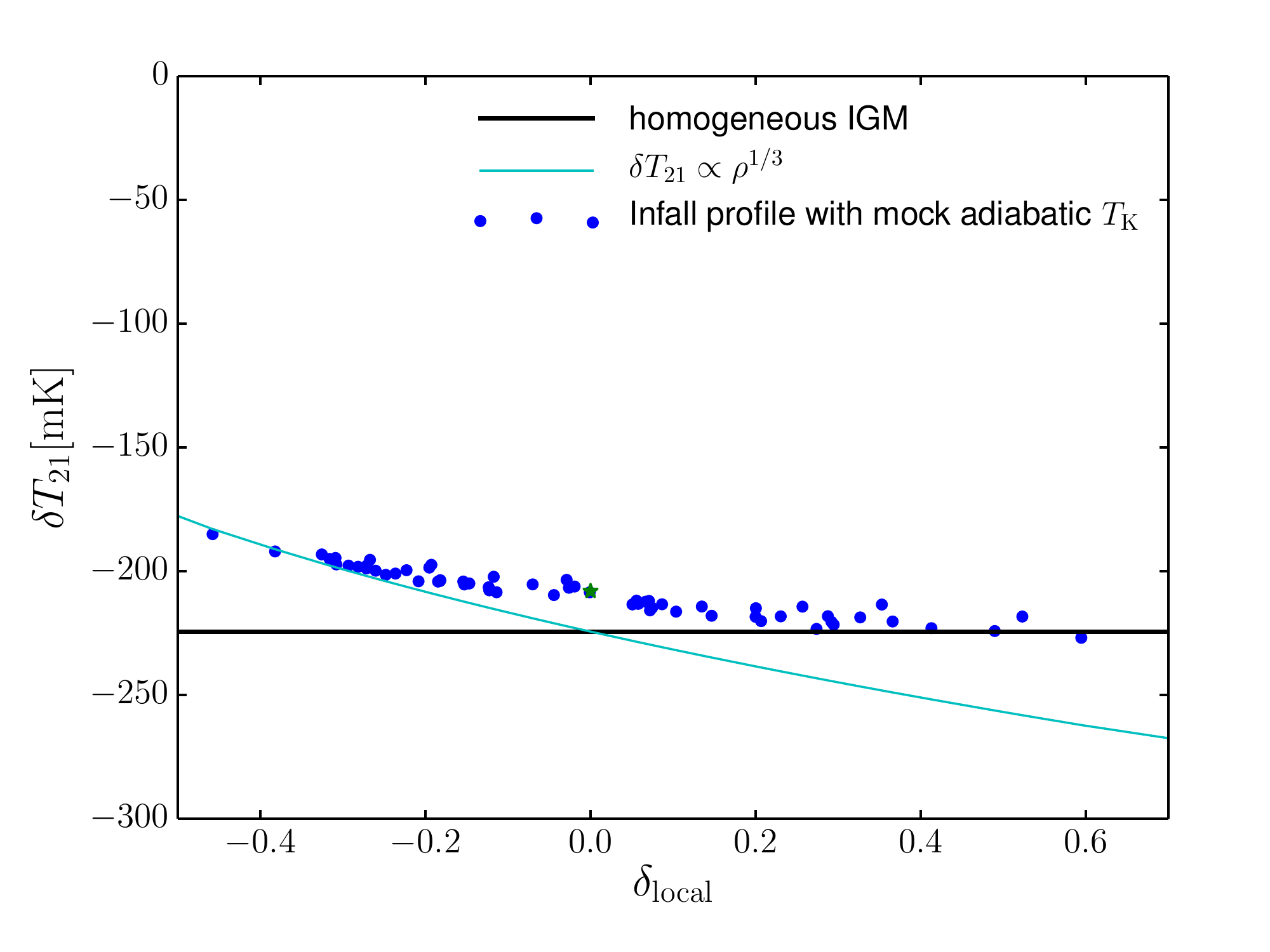}
\includegraphics[scale=0.38]{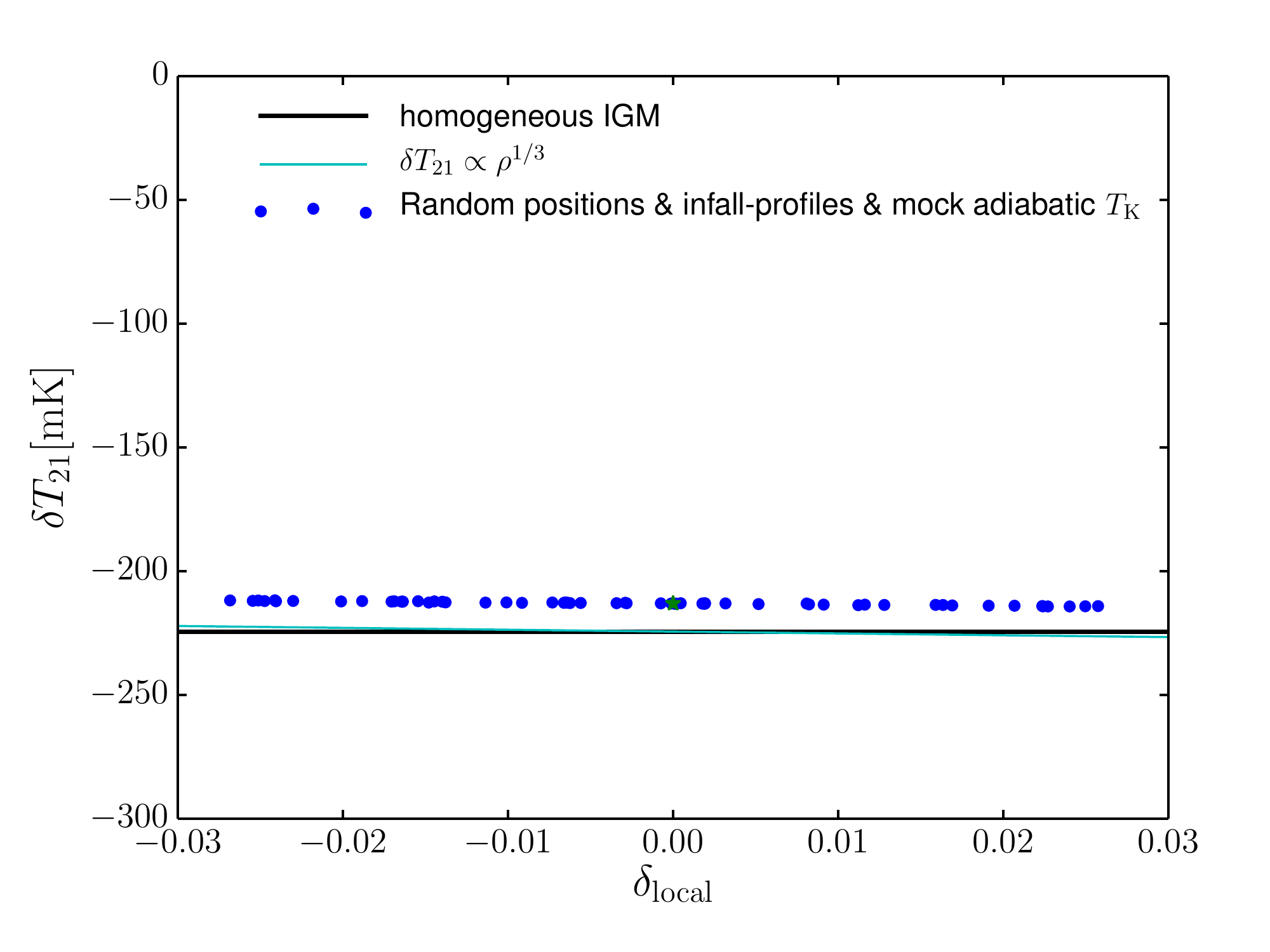}
\caption{The mean 21 cm brightness temperature of sub-boxes with different mean overdensity $\delta_{\rm local}$ 
at  $z = 17$. The mock simulation has a box size of 4 Mpc$/h$ and a resolution of $800^3$. 
{\it Left panel:} Computed with the halo catalog (masses and positions) from the hydrodynamic simulation; 
{\it Right panel:}  Computed with the same halo mass catalog but with randomized halo positions.
Both panels assume the density profiles of the IGM predicted by the infall model.  
The black line indicates the expected value for a homogeneous IGM, and the cyan line shows the 
scaling of $\delta T_{\rm 21} \propto \rho^{1/3}$ for cold optical-thin gas.}
\label{Fig.clustering}
}
\end{figure*}

The large-scale clustering generated during the structure formation may also affect the global 21 cm signal. 
We investigate the effect of clustering by comparing two mock simulations; one using the halo catalog with both
mass and position information from the hydrodynamic simulation (``Mock-clustering'' simulation), 
and the other using only the halo mass catalog with random halo positions (``Mock-random'' simulation). 
Both use the infall model with appropriate normalization to predict the gas density distribution in the IGM, 
and the mock adiabatic temperature is adopted.
The mean 21 cm brightness temperatures of the 64 sub-boxes are plotted
in the left panel for the ``Mock-clustering'' simulation and in the right panel for the ``Mock-random'' simulation,
respectively in Fig.~\ref{Fig.clustering}.
Note the range of density fluctuations $\delta_{\rm local}$ are much smaller for the ``Mock-random'' case.
With the clustered positions of halos, the averaged 21 cm brightness temperature is 
$-208 \mK$ (the green star in the plot), 
which is about 7\% decrement in the absolute value with respect to the homogeneous IGM case,   
while if the halos are randomly distributed, then 
$\delta T_{\rm 21} \sim -213 \mK$, which is only about 5\% decrement in the signal level.   
Therefore, the clustering effect also reduces the absorption level of the 21 cm signal, 
by introducing higher level of density fluctuations, but the effect is only moderate.

\section{Conclusions and discussions}\label{conclusions}

In this work, we investigate the maximum signal level of the global 21 cm spectrum from cosmic dawn 
that could be achieved in the standard cosmology, and discuss various theoretical effects that could have impacts on
the absorption level. 
By running a set of high resolution hydrodynamic simulations, we find that
the non-linear structure formation affects the IGM density and temperature distribution significantly.
The shock-heating and Compton-heating, the non-linear density fluctuations, and the halo clustering, all have non-negligible
effects reducing the 21 cm absorption signal. 
Under the assumption of saturated coupling between the spin temperature of hydrogen
and the gas temperature, the maximum absorption level that is achievable in the standard framework is 
reduce by about 15\% at redshift 17, as compare to the homogeneous IGM case.

Among the various effects considered here, the shock-heating during the non-linear structure formation and the Compton-heating 
play a dominant role in reducing the maximum absorption level. The non-linear density fluctuations with
adiabatic heating can also reduce the contribution from over-dense regions, but the effect is moderate.
The clustering of halos, on the other hand, also enhances the density fluctuations and reduces the 21 cm signal
mildly. By comparing the density profiles in the simulated IGM and those predicted by the infall model,  
we find that the infall model provides a fairly reasonable prediction
for the density distribution around density peaks in the IGM.

We note that the heating effect of structure formation shocks during the cosmic dawn is still somewhat uncertain. 
The early work by \citet{2004ApJ...608..611G} shows that the shock heating has a dramatic effect, 
dominating over Ly-$\alpha$ heating and X-ray heating at high redshifts, and reduces the 21 cm global absorption substantially.
Nevertheless, latter works \citep{2004ApJ...611..642F,2006PhR...433..181F,McQuinn2012} show that the structure formation shocks have only modest effect in heating the gas, being subdominant to X-rays, though the relative importance depends on the timing of the X-ray heating. Even if we disregard the uncertainties in the X-ray production, there are still significant theoretical uncertainties on the structure formation shocks. To capture shocks, 
the SPH algorithm, which is adopted in the GADGET-2 used here, 
introduces an artificial viscosity to provide the entropy generated by microphysics process. 
This introduces unphysical extra heating and broadens the shock front \citep{SPH1992,Gadget2}. 
It may lead to the overcooling problem \citep{Creasey2011,Nelson2013}, and produce artificial cold blobs near the star-forming regions \citep{Hobbs2013}. \citet{Oshea2005} did find that the gas properties in the SPH and the adaptive mesh refinement (AMR) based 
simulations generally agree with each other. In particular, \citet{McQuinn2012} found that for the same resolution, 
the GADGET-2 and the AMR-based Enzo code give quite similar evolution of the mean gas temperature. 
Our results are all based on the GADGET-2 simulations, and they provide reasonable predictions for the scales 
we are interested in. Nevertheless,  \citet{Jia2020} noted that there are still significant divergence in the number and strength of structure formation shocks among different numerical schemes, such uncertainties could affect the results obtained here.

Throughout our calculation, we have assumed that the IGM is totally neutral during the cosmic dawn, in order to estimate the maximum absorption level.
However, we note that the first star formation has to occur in massive halos at rare density peaks, so as to provide the necessary Ly-$\alpha$ photons for spin temperature coupling.
In order to assess the effect of ionization in the densest regions, we select all possibly star-forming halos with a virial temperature 
threshold of $10^4 \K$, and 16 star-forming halos are identified at redshift 17 in our fiducial simulation. 
Then the evolution of the radii of ionized 
bubbles around these individual star-forming halos can be computed (see, e.g. \citealt{2011MNRAS.410.2025X}).
We assume a Salpeter initial mass function (with a power-law slope of $\alpha = 2.35$) for the first stars with 
a mass range of 1 – 500 ${\rm M_\odot}$, and a fixed metallicity of $Z = 10^{-7}$, then the emission rate of ionizing photons
can be determined from the ionizing continua of high-redshift starburst galaxies\footnote{Data are available at http://cdsarc.u-strasbg.fr/cgi-bin/Cat?VI/109.} \citep{Schaerer2002,Schaerer2003}.
Adopting a clumping factor of 3 – 5 as appropriate for high redshift IGM (e.g. \citealt{2014ApJ...787..146K,2020MNRAS.491.1600M}), 
a star formation efficiency of $f_\star = 0.01$, an escape fraction of $f_{\rm esc} = 0.07$, we obtain a global ionized fraction of about 1\%. 
Eliminating all the 21 cm signal from these ionized regions result in mild suppression of the absorption signal, 
with $\delta T_{\rm 21}\approx -189 \mK$ at $z = 17$.
Note that the parameters for the first star formation are quite uncertain, and regardless of these uncertain parameters, the maximum 21 cm absorption signal is achieved if exactly the emission regions are ionized around the star-forming halos, 
and this signal is found to be still about $-190 \mK$, very close to the maximum signal level if the IGM is totally neutral.
Therefore, the emission signal from these very over-dense regions is almost negligible as expected.

In the present work, we have focused on the maximum absorption level, and include only the Compton-heating and 
shock-heating effects that are inevitable during the cosmic dawn, 
isolating them from any other astrophysical heating related to radiation sources.
Note that as more and more galaxies form, various feedback effects including photoionization and X-ray heating 
would gradually dominates over the shock-heating and Compton-heating, the non-linear density fluctuations, 
and the clustering effects, reducing the global 21 cm absorption more significantly.  
Our results provide a modified signal base for any other feedback processes to take further effects on, and indicate that 
one has to take into account the effects of non-linear structure formation in order to accurately interpret 
upcoming observational data, and/or inferring any requirement of new physics (e.g. \citealt{2018Natur.555...71B,2020PhRvD.102h3538Y}).

\acknowledgments
We thank Anastasia Fialkov, Rennan Barkana, Joe Silk, and Paul R. Shapiro for helpful discussions.
This work is supported by National SKA Program of China No. 2020SKA0110401, 
the National Natural Science Foundation of China (NSFC) grant 11973047, 11633004, 
the Chinese Academy of Sciences (CAS) Strategic Priority Research Program XDA15020200, 
the MoST-BRICS Flagship Project No. 2018YFE0120800, 
the CAS Frontier Science Key Project QYZDJ-SSW-SLH017, and
the NSFC-ISF joint research program No. 11761141012. 
BY acknowledges the support of  the BR program from the 
CAS, the NSFC grant 11653003 and the NSFC-CAS  joint fund for space scientific satellites No. U1738125.
This work used resources of China SKA Regional Centre prototype \citep{2019NatAs...3.1030A} funded by the National Key R\&D Programme of China (2018YFA0404603) and Chinese Academy of Sciences (114231KYSB20170003).

\bibliography{references}
\bibliographystyle{hapj}

\end{document}